\documentstyle[psfig,epsfig,color]{mn}

\newif\ifAMStwofonts

\definecolor{AliceBlue}{rgb}{0.94,0.97,1.00}
\definecolor{AntiqueWhite1}{rgb}{1.00,0.94,0.86}
\definecolor{AntiqueWhite2}{rgb}{0.93,0.87,0.80}
\definecolor{AntiqueWhite3}{rgb}{0.80,0.75,0.69}
\definecolor{AntiqueWhite4}{rgb}{0.55,0.51,0.47}
\definecolor{AntiqueWhite}{rgb}{0.98,0.92,0.84}
\definecolor{BlanchedAlmond}{rgb}{1.00,0.92,0.80}
\definecolor{BlueViolet}{rgb}{0.54,0.17,0.89}
\definecolor{CadetBlue1}{rgb}{0.60,0.96,1.00}
\definecolor{CadetBlue2}{rgb}{0.56,0.90,0.93}
\definecolor{CadetBlue3}{rgb}{0.48,0.77,0.80}
\definecolor{CadetBlue4}{rgb}{0.33,0.53,0.55}
\definecolor{CadetBlue}{rgb}{0.37,0.62,0.63}
\definecolor{CornflowerBlue}{rgb}{0.39,0.58,0.93}
\definecolor{DarkBlue}{rgb}{0.00,0.00,0.55}
\definecolor{DarkCyan}{rgb}{0.00,0.55,0.55}
\definecolor{DarkGoldenrod1}{rgb}{1.00,0.73,0.06}
\definecolor{DarkGoldenrod2}{rgb}{0.93,0.68,0.05}
\definecolor{DarkGoldenrod3}{rgb}{0.80,0.58,0.05}
\definecolor{DarkGoldenrod4}{rgb}{0.55,0.40,0.03}
\definecolor{DarkGoldenrod}{rgb}{0.72,0.53,0.04}
\definecolor{DarkGray}{rgb}{0.66,0.66,0.66}
\definecolor{DarkGreen}{rgb}{0.00,0.39,0.00}
\definecolor{DarkGrey}{rgb}{0.66,0.66,0.66}
\definecolor{DarkKhaki}{rgb}{0.74,0.72,0.42}
\definecolor{DarkMagenta}{rgb}{0.55,0.00,0.55}
\definecolor{DarkOliveGreen1}{rgb}{0.79,1.00,0.44}
\definecolor{DarkOliveGreen2}{rgb}{0.74,0.93,0.41}
\definecolor{DarkOliveGreen3}{rgb}{0.64,0.80,0.35}
\definecolor{DarkOliveGreen4}{rgb}{0.43,0.55,0.24}
\definecolor{DarkOliveGreen}{rgb}{0.33,0.42,0.18}
\definecolor{DarkOrange1}{rgb}{1.00,0.50,0.00}
\definecolor{DarkOrange2}{rgb}{0.93,0.46,0.00}
\definecolor{DarkOrange3}{rgb}{0.80,0.40,0.00}
\definecolor{DarkOrange4}{rgb}{0.55,0.27,0.00}
\definecolor{DarkOrange}{rgb}{1.00,0.55,0.00}
\definecolor{DarkOrchid1}{rgb}{0.75,0.24,1.00}
\definecolor{DarkOrchid2}{rgb}{0.70,0.23,0.93}
\definecolor{DarkOrchid3}{rgb}{0.60,0.20,0.80}
\definecolor{DarkOrchid4}{rgb}{0.41,0.13,0.55}
\definecolor{DarkOrchid}{rgb}{0.60,0.20,0.80}
\definecolor{DarkRed}{rgb}{0.55,0.00,0.00}
\definecolor{DarkSalmon}{rgb}{0.91,0.59,0.48}
\definecolor{DarkSeaGreen1}{rgb}{0.76,1.00,0.76}
\definecolor{DarkSeaGreen2}{rgb}{0.71,0.93,0.71}
\definecolor{DarkSeaGreen3}{rgb}{0.61,0.80,0.61}
\definecolor{DarkSeaGreen4}{rgb}{0.41,0.55,0.41}
\definecolor{DarkSeaGreen}{rgb}{0.56,0.74,0.56}
\definecolor{DarkSlateBlue}{rgb}{0.28,0.24,0.55}
\definecolor{DarkSlateGray1}{rgb}{0.59,1.00,1.00}
\definecolor{DarkSlateGray2}{rgb}{0.55,0.93,0.93}
\definecolor{DarkSlateGray3}{rgb}{0.47,0.80,0.80}
\definecolor{DarkSlateGray4}{rgb}{0.32,0.55,0.55}
\definecolor{DarkSlateGray}{rgb}{0.18,0.31,0.31}
\definecolor{DarkSlateGrey}{rgb}{0.18,0.31,0.31}
\definecolor{DarkTurquoise}{rgb}{0.00,0.81,0.82}
\definecolor{DarkViolet}{rgb}{0.58,0.00,0.83}
\definecolor{DeepPink1}{rgb}{1.00,0.08,0.58}
\definecolor{DeepPink2}{rgb}{0.93,0.07,0.54}
\definecolor{DeepPink3}{rgb}{0.80,0.06,0.46}
\definecolor{DeepPink4}{rgb}{0.55,0.04,0.31}
\definecolor{DeepPink}{rgb}{1.00,0.08,0.58}
\definecolor{DeepSkyBlue1}{rgb}{0.00,0.75,1.00}
\definecolor{DeepSkyBlue2}{rgb}{0.00,0.70,0.93}
\definecolor{DeepSkyBlue3}{rgb}{0.00,0.60,0.80}
\definecolor{DeepSkyBlue4}{rgb}{0.00,0.41,0.55}
\definecolor{DeepSkyBlue}{rgb}{0.00,0.75,1.00}
\definecolor{DimGray}{rgb}{0.41,0.41,0.41}
\definecolor{DimGrey}{rgb}{0.41,0.41,0.41}
\definecolor{DodgerBlue1}{rgb}{0.12,0.56,1.00}
\definecolor{DodgerBlue2}{rgb}{0.11,0.53,0.93}
\definecolor{DodgerBlue3}{rgb}{0.09,0.45,0.80}
\definecolor{DodgerBlue4}{rgb}{0.06,0.31,0.55}
\definecolor{DodgerBlue}{rgb}{0.12,0.56,1.00}
\definecolor{FloralWhite}{rgb}{1.00,0.98,0.94}
\definecolor{ForestGreen}{rgb}{0.13,0.55,0.13}
\definecolor{GhostWhite}{rgb}{0.97,0.97,1.00}
\definecolor{GreenYellow}{rgb}{0.68,1.00,0.18}
\definecolor{HotPink1}{rgb}{1.00,0.43,0.71}
\definecolor{HotPink2}{rgb}{0.93,0.42,0.65}
\definecolor{HotPink3}{rgb}{0.80,0.38,0.56}
\definecolor{HotPink4}{rgb}{0.55,0.23,0.38}
\definecolor{HotPink}{rgb}{1.00,0.41,0.71}
\definecolor{IndianRed1}{rgb}{1.00,0.42,0.42}
\definecolor{IndianRed2}{rgb}{0.93,0.39,0.39}
\definecolor{IndianRed3}{rgb}{0.80,0.33,0.33}
\definecolor{IndianRed4}{rgb}{0.55,0.23,0.23}
\definecolor{IndianRed}{rgb}{0.80,0.36,0.36}
\definecolor{LavenderBlush1}{rgb}{1.00,0.94,0.96}
\definecolor{LavenderBlush2}{rgb}{0.93,0.88,0.90}
\definecolor{LavenderBlush3}{rgb}{0.80,0.76,0.77}
\definecolor{LavenderBlush4}{rgb}{0.55,0.51,0.53}
\definecolor{LavenderBlush}{rgb}{1.00,0.94,0.96}
\definecolor{LawnGreen}{rgb}{0.49,0.99,0.00}
\definecolor{LemonChiffon1}{rgb}{1.00,0.98,0.80}
\definecolor{LemonChiffon2}{rgb}{0.93,0.91,0.75}
\definecolor{LemonChiffon3}{rgb}{0.80,0.79,0.65}
\definecolor{LemonChiffon4}{rgb}{0.55,0.54,0.44}
\definecolor{LemonChiffon}{rgb}{1.00,0.98,0.80}
\definecolor{LightBlue1}{rgb}{0.75,0.94,1.00}
\definecolor{LightBlue2}{rgb}{0.70,0.87,0.93}
\definecolor{LightBlue3}{rgb}{0.60,0.75,0.80}
\definecolor{LightBlue4}{rgb}{0.41,0.51,0.55}
\definecolor{LightBlue}{rgb}{0.68,0.85,0.90}
\definecolor{LightCoral}{rgb}{0.94,0.50,0.50}
\definecolor{LightCyan1}{rgb}{0.88,1.00,1.00}
\definecolor{LightCyan2}{rgb}{0.82,0.93,0.93}
\definecolor{LightCyan3}{rgb}{0.71,0.80,0.80}
\definecolor{LightCyan4}{rgb}{0.48,0.55,0.55}
\definecolor{LightCyan}{rgb}{0.88,1.00,1.00}
\definecolor{LightGoldenrod1}{rgb}{1.00,0.93,0.55}
\definecolor{LightGoldenrod2}{rgb}{0.93,0.86,0.51}
\definecolor{LightGoldenrod3}{rgb}{0.80,0.75,0.44}
\definecolor{LightGoldenrod4}{rgb}{0.55,0.51,0.30}
\definecolor{LightGoldenrodYellow}{rgb}{0.98,0.98,0.82}
\definecolor{LightGoldenrod}{rgb}{0.93,0.87,0.51}
\definecolor{LightGray}{rgb}{0.83,0.83,0.83}
\definecolor{LightGreen}{rgb}{0.56,0.93,0.56}
\definecolor{LightGrey}{rgb}{0.83,0.83,0.83}
\definecolor{LightPink1}{rgb}{1.00,0.68,0.73}
\definecolor{LightPink2}{rgb}{0.93,0.64,0.68}
\definecolor{LightPink3}{rgb}{0.80,0.55,0.58}
\definecolor{LightPink4}{rgb}{0.55,0.37,0.40}
\definecolor{LightPink}{rgb}{1.00,0.71,0.76}
\definecolor{LightSalmon1}{rgb}{1.00,0.63,0.48}
\definecolor{LightSalmon2}{rgb}{0.93,0.58,0.45}
\definecolor{LightSalmon3}{rgb}{0.80,0.51,0.38}
\definecolor{LightSalmon4}{rgb}{0.55,0.34,0.26}
\definecolor{LightSalmon}{rgb}{1.00,0.63,0.48}
\definecolor{LightSeaGreen}{rgb}{0.13,0.70,0.67}
\definecolor{LightSkyBlue1}{rgb}{0.69,0.89,1.00}
\definecolor{LightSkyBlue2}{rgb}{0.64,0.83,0.93}
\definecolor{LightSkyBlue3}{rgb}{0.55,0.71,0.80}
\definecolor{LightSkyBlue4}{rgb}{0.38,0.48,0.55}
\definecolor{LightSkyBlue}{rgb}{0.53,0.81,0.98}
\definecolor{LightSlateBlue}{rgb}{0.52,0.44,1.00}
\definecolor{LightSlateGray}{rgb}{0.47,0.53,0.60}
\definecolor{LightSlateGrey}{rgb}{0.47,0.53,0.60}
\definecolor{LightSteelBlue1}{rgb}{0.79,0.88,1.00}
\definecolor{LightSteelBlue2}{rgb}{0.74,0.82,0.93}
\definecolor{LightSteelBlue3}{rgb}{0.64,0.71,0.80}
\definecolor{LightSteelBlue4}{rgb}{0.43,0.48,0.55}
\definecolor{LightSteelBlue}{rgb}{0.69,0.77,0.87}
\definecolor{LightYellow1}{rgb}{1.00,1.00,0.88}
\definecolor{LightYellow2}{rgb}{0.93,0.93,0.82}
\definecolor{LightYellow3}{rgb}{0.80,0.80,0.71}
\definecolor{LightYellow4}{rgb}{0.55,0.55,0.48}
\definecolor{LightYellow}{rgb}{1.00,1.00,0.88}
\definecolor{LimeGreen}{rgb}{0.20,0.80,0.20}
\definecolor{MediumAquamarine}{rgb}{0.40,0.80,0.67}
\definecolor{MediumBlue}{rgb}{0.00,0.00,0.80}
\definecolor{MediumOrchid1}{rgb}{0.88,0.40,1.00}
\definecolor{MediumOrchid2}{rgb}{0.82,0.37,0.93}
\definecolor{MediumOrchid3}{rgb}{0.71,0.32,0.80}
\definecolor{MediumOrchid4}{rgb}{0.48,0.22,0.55}
\definecolor{MediumOrchid}{rgb}{0.73,0.33,0.83}
\definecolor{MediumPurple1}{rgb}{0.67,0.51,1.00}
\definecolor{MediumPurple2}{rgb}{0.62,0.47,0.93}
\definecolor{MediumPurple3}{rgb}{0.54,0.41,0.80}
\definecolor{MediumPurple4}{rgb}{0.36,0.28,0.55}
\definecolor{MediumPurple}{rgb}{0.58,0.44,0.86}
\definecolor{MediumSeaGreen}{rgb}{0.24,0.70,0.44}
\definecolor{MediumSlateBlue}{rgb}{0.48,0.41,0.93}
\definecolor{MediumSpringGreen}{rgb}{0.00,0.98,0.60}
\definecolor{MediumTurquoise}{rgb}{0.28,0.82,0.80}
\definecolor{MediumVioletRed}{rgb}{0.78,0.08,0.52}
\definecolor{MidnightBlue}{rgb}{0.10,0.10,0.44}
\definecolor{MintCream}{rgb}{0.96,1.00,0.98}
\definecolor{MistyRose1}{rgb}{1.00,0.89,0.88}
\definecolor{MistyRose2}{rgb}{0.93,0.84,0.82}
\definecolor{MistyRose3}{rgb}{0.80,0.72,0.71}
\definecolor{MistyRose4}{rgb}{0.55,0.49,0.48}
\definecolor{MistyRose}{rgb}{1.00,0.89,0.88}
\definecolor{NavajoWhite1}{rgb}{1.00,0.87,0.68}
\definecolor{NavajoWhite2}{rgb}{0.93,0.81,0.63}
\definecolor{NavajoWhite3}{rgb}{0.80,0.70,0.55}
\definecolor{NavajoWhite4}{rgb}{0.55,0.47,0.37}
\definecolor{NavajoWhite}{rgb}{1.00,0.87,0.68}
\definecolor{NavyBlue}{rgb}{0.00,0.00,0.50}
\definecolor{OldLace}{rgb}{0.99,0.96,0.90}
\definecolor{OliveDrab1}{rgb}{0.75,1.00,0.24}
\definecolor{OliveDrab2}{rgb}{0.70,0.93,0.23}
\definecolor{OliveDrab3}{rgb}{0.60,0.80,0.20}
\definecolor{OliveDrab4}{rgb}{0.41,0.55,0.13}
\definecolor{OliveDrab}{rgb}{0.42,0.56,0.14}
\definecolor{OrangeRed1}{rgb}{1.00,0.27,0.00}
\definecolor{OrangeRed2}{rgb}{0.93,0.25,0.00}
\definecolor{OrangeRed3}{rgb}{0.80,0.22,0.00}
\definecolor{OrangeRed4}{rgb}{0.55,0.15,0.00}
\definecolor{OrangeRed}{rgb}{1.00,0.27,0.00}
\definecolor{PaleGoldenrod}{rgb}{0.93,0.91,0.67}
\definecolor{PaleGreen1}{rgb}{0.60,1.00,0.60}
\definecolor{PaleGreen2}{rgb}{0.56,0.93,0.56}
\definecolor{PaleGreen3}{rgb}{0.49,0.80,0.49}
\definecolor{PaleGreen4}{rgb}{0.33,0.55,0.33}
\definecolor{PaleGreen}{rgb}{0.60,0.98,0.60}
\definecolor{PaleTurquoise1}{rgb}{0.73,1.00,1.00}
\definecolor{PaleTurquoise2}{rgb}{0.68,0.93,0.93}
\definecolor{PaleTurquoise3}{rgb}{0.59,0.80,0.80}
\definecolor{PaleTurquoise4}{rgb}{0.40,0.55,0.55}
\definecolor{PaleTurquoise}{rgb}{0.69,0.93,0.93}
\definecolor{PaleVioletRed1}{rgb}{1.00,0.51,0.67}
\definecolor{PaleVioletRed2}{rgb}{0.93,0.47,0.62}
\definecolor{PaleVioletRed3}{rgb}{0.80,0.41,0.54}
\definecolor{PaleVioletRed4}{rgb}{0.55,0.28,0.36}
\definecolor{PaleVioletRed}{rgb}{0.86,0.44,0.58}
\definecolor{PapayaWhip}{rgb}{1.00,0.94,0.84}
\definecolor{PeachPuff1}{rgb}{1.00,0.85,0.73}
\definecolor{PeachPuff2}{rgb}{0.93,0.80,0.68}
\definecolor{PeachPuff3}{rgb}{0.80,0.69,0.58}
\definecolor{PeachPuff4}{rgb}{0.55,0.47,0.40}
\definecolor{PeachPuff}{rgb}{1.00,0.85,0.73}
\definecolor{PowderBlue}{rgb}{0.69,0.88,0.90}
\definecolor{RosyBrown1}{rgb}{1.00,0.76,0.76}
\definecolor{RosyBrown2}{rgb}{0.93,0.71,0.71}
\definecolor{RosyBrown3}{rgb}{0.80,0.61,0.61}
\definecolor{RosyBrown4}{rgb}{0.55,0.41,0.41}
\definecolor{RosyBrown}{rgb}{0.74,0.56,0.56}
\definecolor{RoyalBlue1}{rgb}{0.28,0.46,1.00}
\definecolor{RoyalBlue2}{rgb}{0.26,0.43,0.93}
\definecolor{RoyalBlue3}{rgb}{0.23,0.37,0.80}
\definecolor{RoyalBlue4}{rgb}{0.15,0.25,0.55}
\definecolor{RoyalBlue}{rgb}{0.25,0.41,0.88}
\definecolor{SaddleBrown}{rgb}{0.55,0.27,0.07}
\definecolor{SandyBrown}{rgb}{0.96,0.64,0.38}
\definecolor{SeaGreen1}{rgb}{0.33,1.00,0.62}
\definecolor{SeaGreen2}{rgb}{0.31,0.93,0.58}
\definecolor{SeaGreen3}{rgb}{0.26,0.80,0.50}
\definecolor{SeaGreen4}{rgb}{0.18,0.55,0.34}
\definecolor{SeaGreen}{rgb}{0.18,0.55,0.34}
\definecolor{SkyBlue1}{rgb}{0.53,0.81,1.00}
\definecolor{SkyBlue2}{rgb}{0.49,0.75,0.93}
\definecolor{SkyBlue3}{rgb}{0.42,0.65,0.80}
\definecolor{SkyBlue4}{rgb}{0.29,0.44,0.55}
\definecolor{SkyBlue}{rgb}{0.53,0.81,0.92}
\definecolor{SlateBlue1}{rgb}{0.51,0.44,1.00}
\definecolor{SlateBlue2}{rgb}{0.48,0.40,0.93}
\definecolor{SlateBlue3}{rgb}{0.41,0.35,0.80}
\definecolor{SlateBlue4}{rgb}{0.28,0.24,0.55}
\definecolor{SlateBlue}{rgb}{0.42,0.35,0.80}
\definecolor{SlateGray1}{rgb}{0.78,0.89,1.00}
\definecolor{SlateGray2}{rgb}{0.73,0.83,0.93}
\definecolor{SlateGray3}{rgb}{0.62,0.71,0.80}
\definecolor{SlateGray4}{rgb}{0.42,0.48,0.55}
\definecolor{SlateGray}{rgb}{0.44,0.50,0.56}
\definecolor{SlateGrey}{rgb}{0.44,0.50,0.56}
\definecolor{SpringGreen1}{rgb}{0.00,1.00,0.50}
\definecolor{SpringGreen2}{rgb}{0.00,0.93,0.46}
\definecolor{SpringGreen3}{rgb}{0.00,0.80,0.40}
\definecolor{SpringGreen4}{rgb}{0.00,0.55,0.27}
\definecolor{SpringGreen}{rgb}{0.00,1.00,0.50}
\definecolor{SteelBlue1}{rgb}{0.39,0.72,1.00}
\definecolor{SteelBlue2}{rgb}{0.36,0.67,0.93}
\definecolor{SteelBlue3}{rgb}{0.31,0.58,0.80}
\definecolor{SteelBlue4}{rgb}{0.21,0.39,0.55}
\definecolor{SteelBlue}{rgb}{0.27,0.51,0.71}
\definecolor{VioletRed1}{rgb}{1.00,0.24,0.59}
\definecolor{VioletRed2}{rgb}{0.93,0.23,0.55}
\definecolor{VioletRed3}{rgb}{0.80,0.20,0.47}
\definecolor{VioletRed4}{rgb}{0.55,0.13,0.32}
\definecolor{VioletRed}{rgb}{0.82,0.13,0.56}
\definecolor{WhiteSmoke}{rgb}{0.96,0.96,0.96}
\definecolor{YellowGreen}{rgb}{0.60,0.80,0.20}
\definecolor{aliceblue}{rgb}{0.94,0.97,1.00}
\definecolor{antiquewhite}{rgb}{0.98,0.92,0.84}
\definecolor{aquamarine1}{rgb}{0.50,1.00,0.83}
\definecolor{aquamarine2}{rgb}{0.46,0.93,0.78}
\definecolor{aquamarine3}{rgb}{0.40,0.80,0.67}
\definecolor{aquamarine4}{rgb}{0.27,0.55,0.45}
\definecolor{aquamarine}{rgb}{0.50,1.00,0.83}
\definecolor{azure1}{rgb}{0.94,1.00,1.00}
\definecolor{azure2}{rgb}{0.88,0.93,0.93}
\definecolor{azure3}{rgb}{0.76,0.80,0.80}
\definecolor{azure4}{rgb}{0.51,0.55,0.55}
\definecolor{azure}{rgb}{0.94,1.00,1.00}
\definecolor{beige}{rgb}{0.96,0.96,0.86}
\definecolor{bisque1}{rgb}{1.00,0.89,0.77}
\definecolor{bisque2}{rgb}{0.93,0.84,0.72}
\definecolor{bisque3}{rgb}{0.80,0.72,0.62}
\definecolor{bisque4}{rgb}{0.55,0.49,0.42}
\definecolor{bisque}{rgb}{1.00,0.89,0.77}
\definecolor{black}{rgb}{0.00,0.00,0.00}
\definecolor{blanchedalmond}{rgb}{1.00,0.92,0.80}
\definecolor{blue1}{rgb}{0.00,0.00,1.00}
\definecolor{blue2}{rgb}{0.00,0.00,0.93}
\definecolor{blue3}{rgb}{0.00,0.00,0.80}
\definecolor{blue4}{rgb}{0.00,0.00,0.55}
\definecolor{blueviolet}{rgb}{0.54,0.17,0.89}
\definecolor{blue}{rgb}{0.00,0.00,1.00}
\definecolor{brown1}{rgb}{1.00,0.25,0.25}
\definecolor{brown2}{rgb}{0.93,0.23,0.23}
\definecolor{brown3}{rgb}{0.80,0.20,0.20}
\definecolor{brown4}{rgb}{0.55,0.14,0.14}
\definecolor{brown}{rgb}{0.65,0.16,0.16}
\definecolor{burlywood1}{rgb}{1.00,0.83,0.61}
\definecolor{burlywood2}{rgb}{0.93,0.77,0.57}
\definecolor{burlywood3}{rgb}{0.80,0.67,0.49}
\definecolor{burlywood4}{rgb}{0.55,0.45,0.33}
\definecolor{burlywood}{rgb}{0.87,0.72,0.53}
\definecolor{cadetblue}{rgb}{0.37,0.62,0.63}
\definecolor{chartreuse1}{rgb}{0.50,1.00,0.00}
\definecolor{chartreuse2}{rgb}{0.46,0.93,0.00}
\definecolor{chartreuse3}{rgb}{0.40,0.80,0.00}
\definecolor{chartreuse4}{rgb}{0.27,0.55,0.00}
\definecolor{chartreuse}{rgb}{0.50,1.00,0.00}
\definecolor{chocolate1}{rgb}{1.00,0.50,0.14}
\definecolor{chocolate2}{rgb}{0.93,0.46,0.13}
\definecolor{chocolate3}{rgb}{0.80,0.40,0.11}
\definecolor{chocolate4}{rgb}{0.55,0.27,0.07}
\definecolor{chocolate}{rgb}{0.82,0.41,0.12}
\definecolor{coral1}{rgb}{1.00,0.45,0.34}
\definecolor{coral2}{rgb}{0.93,0.42,0.31}
\definecolor{coral3}{rgb}{0.80,0.36,0.27}
\definecolor{coral4}{rgb}{0.55,0.24,0.18}
\definecolor{coral}{rgb}{1.00,0.50,0.31}
\definecolor{cornflowerblue}{rgb}{0.39,0.58,0.93}
\definecolor{cornsilk1}{rgb}{1.00,0.97,0.86}
\definecolor{cornsilk2}{rgb}{0.93,0.91,0.80}
\definecolor{cornsilk3}{rgb}{0.80,0.78,0.69}
\definecolor{cornsilk4}{rgb}{0.55,0.53,0.47}
\definecolor{cornsilk}{rgb}{1.00,0.97,0.86}
\definecolor{cyan1}{rgb}{0.00,1.00,1.00}
\definecolor{cyan2}{rgb}{0.00,0.93,0.93}
\definecolor{cyan3}{rgb}{0.00,0.80,0.80}
\definecolor{cyan4}{rgb}{0.00,0.55,0.55}
\definecolor{cyan}{rgb}{0.00,1.00,1.00}
\definecolor{darkblue}{rgb}{0.00,0.00,0.55}
\definecolor{darkcyan}{rgb}{0.00,0.55,0.55}
\definecolor{darkgoldenrod}{rgb}{0.72,0.53,0.04}
\definecolor{darkgray}{rgb}{0.66,0.66,0.66}
\definecolor{darkgreen}{rgb}{0.00,0.39,0.00}
\definecolor{darkgrey}{rgb}{0.66,0.66,0.66}
\definecolor{darkkhaki}{rgb}{0.74,0.72,0.42}
\definecolor{darkmagenta}{rgb}{0.55,0.00,0.55}
\definecolor{darkolive}{rgb}{0.33,0.42,0.18}
\definecolor{darkorange}{rgb}{1.00,0.55,0.00}
\definecolor{darkorchid}{rgb}{0.60,0.20,0.80}
\definecolor{darkred}{rgb}{0.55,0.00,0.00}
\definecolor{darksalmon}{rgb}{0.91,0.59,0.48}
\definecolor{darksea}{rgb}{0.56,0.74,0.56}
\definecolor{darkslate}{rgb}{0.18,0.31,0.31}
\definecolor{darkslate}{rgb}{0.18,0.31,0.31}
\definecolor{darkslate}{rgb}{0.28,0.24,0.55}
\definecolor{darkturquoise}{rgb}{0.00,0.81,0.82}
\definecolor{darkviolet}{rgb}{0.58,0.00,0.83}
\definecolor{deeppink}{rgb}{1.00,0.08,0.58}
\definecolor{deepsky}{rgb}{0.00,0.75,1.00}
\definecolor{dimgray}{rgb}{0.41,0.41,0.41}
\definecolor{dimgrey}{rgb}{0.41,0.41,0.41}
\definecolor{dodgerblue}{rgb}{0.12,0.56,1.00}
\definecolor{firebrick1}{rgb}{1.00,0.19,0.19}
\definecolor{firebrick2}{rgb}{0.93,0.17,0.17}
\definecolor{firebrick3}{rgb}{0.80,0.15,0.15}
\definecolor{firebrick4}{rgb}{0.55,0.10,0.10}
\definecolor{firebrick}{rgb}{0.70,0.13,0.13}
\definecolor{floralwhite}{rgb}{1.00,0.98,0.94}
\definecolor{forestgreen}{rgb}{0.13,0.55,0.13}
\definecolor{gainsboro}{rgb}{0.86,0.86,0.86}
\definecolor{ghostwhite}{rgb}{0.97,0.97,1.00}
\definecolor{gold1}{rgb}{1.00,0.84,0.00}
\definecolor{gold2}{rgb}{0.93,0.79,0.00}
\definecolor{gold3}{rgb}{0.80,0.68,0.00}
\definecolor{gold4}{rgb}{0.55,0.46,0.00}
\definecolor{goldenrod1}{rgb}{1.00,0.76,0.15}
\definecolor{goldenrod2}{rgb}{0.93,0.71,0.13}
\definecolor{goldenrod3}{rgb}{0.80,0.61,0.11}
\definecolor{goldenrod4}{rgb}{0.55,0.41,0.08}
\definecolor{goldenrod}{rgb}{0.85,0.65,0.13}
\definecolor{gold}{rgb}{1.00,0.84,0.00}
\definecolor{gray0}{rgb}{0.00,0.00,0.00}
\definecolor{gray100}{rgb}{1.00,1.00,1.00}
\definecolor{gray10}{rgb}{0.10,0.10,0.10}
\definecolor{gray11}{rgb}{0.11,0.11,0.11}
\definecolor{gray12}{rgb}{0.12,0.12,0.12}
\definecolor{gray13}{rgb}{0.13,0.13,0.13}
\definecolor{gray14}{rgb}{0.14,0.14,0.14}
\definecolor{gray15}{rgb}{0.15,0.15,0.15}
\definecolor{gray16}{rgb}{0.16,0.16,0.16}
\definecolor{gray17}{rgb}{0.17,0.17,0.17}
\definecolor{gray18}{rgb}{0.18,0.18,0.18}
\definecolor{gray19}{rgb}{0.19,0.19,0.19}
\definecolor{gray1}{rgb}{0.01,0.01,0.01}
\definecolor{gray20}{rgb}{0.20,0.20,0.20}
\definecolor{gray21}{rgb}{0.21,0.21,0.21}
\definecolor{gray22}{rgb}{0.22,0.22,0.22}
\definecolor{gray23}{rgb}{0.23,0.23,0.23}
\definecolor{gray24}{rgb}{0.24,0.24,0.24}
\definecolor{gray25}{rgb}{0.25,0.25,0.25}
\definecolor{gray26}{rgb}{0.26,0.26,0.26}
\definecolor{gray27}{rgb}{0.27,0.27,0.27}
\definecolor{gray28}{rgb}{0.28,0.28,0.28}
\definecolor{gray29}{rgb}{0.29,0.29,0.29}
\definecolor{gray2}{rgb}{0.02,0.02,0.02}
\definecolor{gray30}{rgb}{0.30,0.30,0.30}
\definecolor{gray31}{rgb}{0.31,0.31,0.31}
\definecolor{gray32}{rgb}{0.32,0.32,0.32}
\definecolor{gray33}{rgb}{0.33,0.33,0.33}
\definecolor{gray34}{rgb}{0.34,0.34,0.34}
\definecolor{gray35}{rgb}{0.35,0.35,0.35}
\definecolor{gray36}{rgb}{0.36,0.36,0.36}
\definecolor{gray37}{rgb}{0.37,0.37,0.37}
\definecolor{gray38}{rgb}{0.38,0.38,0.38}
\definecolor{gray39}{rgb}{0.39,0.39,0.39}
\definecolor{gray3}{rgb}{0.03,0.03,0.03}
\definecolor{gray40}{rgb}{0.40,0.40,0.40}
\definecolor{gray41}{rgb}{0.41,0.41,0.41}
\definecolor{gray42}{rgb}{0.42,0.42,0.42}
\definecolor{gray43}{rgb}{0.43,0.43,0.43}
\definecolor{gray44}{rgb}{0.44,0.44,0.44}
\definecolor{gray45}{rgb}{0.45,0.45,0.45}
\definecolor{gray46}{rgb}{0.46,0.46,0.46}
\definecolor{gray47}{rgb}{0.47,0.47,0.47}
\definecolor{gray48}{rgb}{0.48,0.48,0.48}
\definecolor{gray49}{rgb}{0.49,0.49,0.49}
\definecolor{gray4}{rgb}{0.04,0.04,0.04}
\definecolor{gray50}{rgb}{0.50,0.50,0.50}
\definecolor{gray51}{rgb}{0.51,0.51,0.51}
\definecolor{gray52}{rgb}{0.52,0.52,0.52}
\definecolor{gray53}{rgb}{0.53,0.53,0.53}
\definecolor{gray54}{rgb}{0.54,0.54,0.54}
\definecolor{gray55}{rgb}{0.55,0.55,0.55}
\definecolor{gray56}{rgb}{0.56,0.56,0.56}
\definecolor{gray57}{rgb}{0.57,0.57,0.57}
\definecolor{gray58}{rgb}{0.58,0.58,0.58}
\definecolor{gray59}{rgb}{0.59,0.59,0.59}
\definecolor{gray5}{rgb}{0.05,0.05,0.05}
\definecolor{gray60}{rgb}{0.60,0.60,0.60}
\definecolor{gray61}{rgb}{0.61,0.61,0.61}
\definecolor{gray62}{rgb}{0.62,0.62,0.62}
\definecolor{gray63}{rgb}{0.63,0.63,0.63}
\definecolor{gray64}{rgb}{0.64,0.64,0.64}
\definecolor{gray65}{rgb}{0.65,0.65,0.65}
\definecolor{gray66}{rgb}{0.66,0.66,0.66}
\definecolor{gray67}{rgb}{0.67,0.67,0.67}
\definecolor{gray68}{rgb}{0.68,0.68,0.68}
\definecolor{gray69}{rgb}{0.69,0.69,0.69}
\definecolor{gray6}{rgb}{0.06,0.06,0.06}
\definecolor{gray70}{rgb}{0.70,0.70,0.70}
\definecolor{gray71}{rgb}{0.71,0.71,0.71}
\definecolor{gray72}{rgb}{0.72,0.72,0.72}
\definecolor{gray73}{rgb}{0.73,0.73,0.73}
\definecolor{gray74}{rgb}{0.74,0.74,0.74}
\definecolor{gray75}{rgb}{0.75,0.75,0.75}
\definecolor{gray76}{rgb}{0.76,0.76,0.76}
\definecolor{gray77}{rgb}{0.77,0.77,0.77}
\definecolor{gray78}{rgb}{0.78,0.78,0.78}
\definecolor{gray79}{rgb}{0.79,0.79,0.79}
\definecolor{gray7}{rgb}{0.07,0.07,0.07}
\definecolor{gray80}{rgb}{0.80,0.80,0.80}
\definecolor{gray81}{rgb}{0.81,0.81,0.81}
\definecolor{gray82}{rgb}{0.82,0.82,0.82}
\definecolor{gray83}{rgb}{0.83,0.83,0.83}
\definecolor{gray84}{rgb}{0.84,0.84,0.84}
\definecolor{gray85}{rgb}{0.85,0.85,0.85}
\definecolor{gray86}{rgb}{0.86,0.86,0.86}
\definecolor{gray87}{rgb}{0.87,0.87,0.87}
\definecolor{gray88}{rgb}{0.88,0.88,0.88}
\definecolor{gray89}{rgb}{0.89,0.89,0.89}
\definecolor{gray8}{rgb}{0.08,0.08,0.08}
\definecolor{gray90}{rgb}{0.90,0.90,0.90}
\definecolor{gray91}{rgb}{0.91,0.91,0.91}
\definecolor{gray92}{rgb}{0.92,0.92,0.92}
\definecolor{gray93}{rgb}{0.93,0.93,0.93}
\definecolor{gray94}{rgb}{0.94,0.94,0.94}
\definecolor{gray95}{rgb}{0.95,0.95,0.95}
\definecolor{gray96}{rgb}{0.96,0.96,0.96}
\definecolor{gray97}{rgb}{0.97,0.97,0.97}
\definecolor{gray98}{rgb}{0.98,0.98,0.98}
\definecolor{gray99}{rgb}{0.99,0.99,0.99}
\definecolor{gray9}{rgb}{0.09,0.09,0.09}
\definecolor{gray}{rgb}{0.75,0.75,0.75}
\definecolor{green1}{rgb}{0.00,1.00,0.00}
\definecolor{green2}{rgb}{0.00,0.93,0.00}
\definecolor{green3}{rgb}{0.00,0.80,0.00}
\definecolor{green4}{rgb}{0.00,0.55,0.00}
\definecolor{greenyellow}{rgb}{0.68,1.00,0.18}
\definecolor{green}{rgb}{0.00,1.00,0.00}
\definecolor{grey0}{rgb}{0.00,0.00,0.00}
\definecolor{grey100}{rgb}{1.00,1.00,1.00}
\definecolor{grey10}{rgb}{0.10,0.10,0.10}
\definecolor{grey11}{rgb}{0.11,0.11,0.11}
\definecolor{grey12}{rgb}{0.12,0.12,0.12}
\definecolor{grey13}{rgb}{0.13,0.13,0.13}
\definecolor{grey14}{rgb}{0.14,0.14,0.14}
\definecolor{grey15}{rgb}{0.15,0.15,0.15}
\definecolor{grey16}{rgb}{0.16,0.16,0.16}
\definecolor{grey17}{rgb}{0.17,0.17,0.17}
\definecolor{grey18}{rgb}{0.18,0.18,0.18}
\definecolor{grey19}{rgb}{0.19,0.19,0.19}
\definecolor{grey1}{rgb}{0.01,0.01,0.01}
\definecolor{grey20}{rgb}{0.20,0.20,0.20}
\definecolor{grey21}{rgb}{0.21,0.21,0.21}
\definecolor{grey22}{rgb}{0.22,0.22,0.22}
\definecolor{grey23}{rgb}{0.23,0.23,0.23}
\definecolor{grey24}{rgb}{0.24,0.24,0.24}
\definecolor{grey25}{rgb}{0.25,0.25,0.25}
\definecolor{grey26}{rgb}{0.26,0.26,0.26}
\definecolor{grey27}{rgb}{0.27,0.27,0.27}
\definecolor{grey28}{rgb}{0.28,0.28,0.28}
\definecolor{grey29}{rgb}{0.29,0.29,0.29}
\definecolor{grey2}{rgb}{0.02,0.02,0.02}
\definecolor{grey30}{rgb}{0.30,0.30,0.30}
\definecolor{grey31}{rgb}{0.31,0.31,0.31}
\definecolor{grey32}{rgb}{0.32,0.32,0.32}
\definecolor{grey33}{rgb}{0.33,0.33,0.33}
\definecolor{grey34}{rgb}{0.34,0.34,0.34}
\definecolor{grey35}{rgb}{0.35,0.35,0.35}
\definecolor{grey36}{rgb}{0.36,0.36,0.36}
\definecolor{grey37}{rgb}{0.37,0.37,0.37}
\definecolor{grey38}{rgb}{0.38,0.38,0.38}
\definecolor{grey39}{rgb}{0.39,0.39,0.39}
\definecolor{grey3}{rgb}{0.03,0.03,0.03}
\definecolor{grey40}{rgb}{0.40,0.40,0.40}
\definecolor{grey41}{rgb}{0.41,0.41,0.41}
\definecolor{grey42}{rgb}{0.42,0.42,0.42}
\definecolor{grey43}{rgb}{0.43,0.43,0.43}
\definecolor{grey44}{rgb}{0.44,0.44,0.44}
\definecolor{grey45}{rgb}{0.45,0.45,0.45}
\definecolor{grey46}{rgb}{0.46,0.46,0.46}
\definecolor{grey47}{rgb}{0.47,0.47,0.47}
\definecolor{grey48}{rgb}{0.48,0.48,0.48}
\definecolor{grey49}{rgb}{0.49,0.49,0.49}
\definecolor{grey4}{rgb}{0.04,0.04,0.04}
\definecolor{grey50}{rgb}{0.50,0.50,0.50}
\definecolor{grey51}{rgb}{0.51,0.51,0.51}
\definecolor{grey52}{rgb}{0.52,0.52,0.52}
\definecolor{grey53}{rgb}{0.53,0.53,0.53}
\definecolor{grey54}{rgb}{0.54,0.54,0.54}
\definecolor{grey55}{rgb}{0.55,0.55,0.55}
\definecolor{grey56}{rgb}{0.56,0.56,0.56}
\definecolor{grey57}{rgb}{0.57,0.57,0.57}
\definecolor{grey58}{rgb}{0.58,0.58,0.58}
\definecolor{grey59}{rgb}{0.59,0.59,0.59}
\definecolor{grey5}{rgb}{0.05,0.05,0.05}
\definecolor{grey60}{rgb}{0.60,0.60,0.60}
\definecolor{grey61}{rgb}{0.61,0.61,0.61}
\definecolor{grey62}{rgb}{0.62,0.62,0.62}
\definecolor{grey63}{rgb}{0.63,0.63,0.63}
\definecolor{grey64}{rgb}{0.64,0.64,0.64}
\definecolor{grey65}{rgb}{0.65,0.65,0.65}
\definecolor{grey66}{rgb}{0.66,0.66,0.66}
\definecolor{grey67}{rgb}{0.67,0.67,0.67}
\definecolor{grey68}{rgb}{0.68,0.68,0.68}
\definecolor{grey69}{rgb}{0.69,0.69,0.69}
\definecolor{grey6}{rgb}{0.06,0.06,0.06}
\definecolor{grey70}{rgb}{0.70,0.70,0.70}
\definecolor{grey71}{rgb}{0.71,0.71,0.71}
\definecolor{grey72}{rgb}{0.72,0.72,0.72}
\definecolor{grey73}{rgb}{0.73,0.73,0.73}
\definecolor{grey74}{rgb}{0.74,0.74,0.74}
\definecolor{grey75}{rgb}{0.75,0.75,0.75}
\definecolor{grey76}{rgb}{0.76,0.76,0.76}
\definecolor{grey77}{rgb}{0.77,0.77,0.77}
\definecolor{grey78}{rgb}{0.78,0.78,0.78}
\definecolor{grey79}{rgb}{0.79,0.79,0.79}
\definecolor{grey7}{rgb}{0.07,0.07,0.07}
\definecolor{grey80}{rgb}{0.80,0.80,0.80}
\definecolor{grey81}{rgb}{0.81,0.81,0.81}
\definecolor{grey82}{rgb}{0.82,0.82,0.82}
\definecolor{grey83}{rgb}{0.83,0.83,0.83}
\definecolor{grey84}{rgb}{0.84,0.84,0.84}
\definecolor{grey85}{rgb}{0.85,0.85,0.85}
\definecolor{grey86}{rgb}{0.86,0.86,0.86}
\definecolor{grey87}{rgb}{0.87,0.87,0.87}
\definecolor{grey88}{rgb}{0.88,0.88,0.88}
\definecolor{grey89}{rgb}{0.89,0.89,0.89}
\definecolor{grey8}{rgb}{0.08,0.08,0.08}
\definecolor{grey90}{rgb}{0.90,0.90,0.90}
\definecolor{grey91}{rgb}{0.91,0.91,0.91}
\definecolor{grey92}{rgb}{0.92,0.92,0.92}
\definecolor{grey93}{rgb}{0.93,0.93,0.93}
\definecolor{grey94}{rgb}{0.94,0.94,0.94}
\definecolor{grey95}{rgb}{0.95,0.95,0.95}
\definecolor{grey96}{rgb}{0.96,0.96,0.96}
\definecolor{grey97}{rgb}{0.97,0.97,0.97}
\definecolor{grey98}{rgb}{0.98,0.98,0.98}
\definecolor{grey99}{rgb}{0.99,0.99,0.99}
\definecolor{grey9}{rgb}{0.09,0.09,0.09}
\definecolor{grey}{rgb}{0.75,0.75,0.75}
\definecolor{honeydew1}{rgb}{0.94,1.00,0.94}
\definecolor{honeydew2}{rgb}{0.88,0.93,0.88}
\definecolor{honeydew3}{rgb}{0.76,0.80,0.76}
\definecolor{honeydew4}{rgb}{0.51,0.55,0.51}
\definecolor{honeydew}{rgb}{0.94,1.00,0.94}
\definecolor{hotpink}{rgb}{1.00,0.41,0.71}
\definecolor{indianred}{rgb}{0.80,0.36,0.36}
\definecolor{ivory1}{rgb}{1.00,1.00,0.94}
\definecolor{ivory2}{rgb}{0.93,0.93,0.88}
\definecolor{ivory3}{rgb}{0.80,0.80,0.76}
\definecolor{ivory4}{rgb}{0.55,0.55,0.51}
\definecolor{ivory}{rgb}{1.00,1.00,0.94}
\definecolor{khaki1}{rgb}{1.00,0.96,0.56}
\definecolor{khaki2}{rgb}{0.93,0.90,0.52}
\definecolor{khaki3}{rgb}{0.80,0.78,0.45}
\definecolor{khaki4}{rgb}{0.55,0.53,0.31}
\definecolor{khaki}{rgb}{0.94,0.90,0.55}
\definecolor{lavenderblush}{rgb}{1.00,0.94,0.96}
\definecolor{lavender}{rgb}{0.90,0.90,0.98}
\definecolor{lawngreen}{rgb}{0.49,0.99,0.00}
\definecolor{lemonchiffon}{rgb}{1.00,0.98,0.80}
\definecolor{lightblue}{rgb}{0.68,0.85,0.90}
\definecolor{lightcoral}{rgb}{0.94,0.50,0.50}
\definecolor{lightcyan}{rgb}{0.88,1.00,1.00}
\definecolor{lightgoldenrod}{rgb}{0.93,0.87,0.51}
\definecolor{lightgoldenrod}{rgb}{0.98,0.98,0.82}
\definecolor{lightgray}{rgb}{0.83,0.83,0.83}
\definecolor{lightgreen}{rgb}{0.56,0.93,0.56}
\definecolor{lightgrey}{rgb}{0.83,0.83,0.83}
\definecolor{lightpink}{rgb}{1.00,0.71,0.76}
\definecolor{lightsalmon}{rgb}{1.00,0.63,0.48}
\definecolor{lightsea}{rgb}{0.13,0.70,0.67}
\definecolor{lightsky}{rgb}{0.53,0.81,0.98}
\definecolor{lightslate}{rgb}{0.47,0.53,0.60}
\definecolor{lightslate}{rgb}{0.47,0.53,0.60}
\definecolor{lightslate}{rgb}{0.52,0.44,1.00}
\definecolor{lightsteel}{rgb}{0.69,0.77,0.87}
\definecolor{lightyellow}{rgb}{1.00,1.00,0.88}
\definecolor{limegreen}{rgb}{0.20,0.80,0.20}
\definecolor{linen}{rgb}{0.98,0.94,0.90}
\definecolor{magenta1}{rgb}{1.00,0.00,1.00}
\definecolor{magenta2}{rgb}{0.93,0.00,0.93}
\definecolor{magenta3}{rgb}{0.80,0.00,0.80}
\definecolor{magenta4}{rgb}{0.55,0.00,0.55}
\definecolor{magenta}{rgb}{1.00,0.00,1.00}
\definecolor{maroon1}{rgb}{1.00,0.20,0.70}
\definecolor{maroon2}{rgb}{0.93,0.19,0.65}
\definecolor{maroon3}{rgb}{0.80,0.16,0.56}
\definecolor{maroon4}{rgb}{0.55,0.11,0.38}
\definecolor{maroon}{rgb}{0.69,0.19,0.38}
\definecolor{mediumaquamarine}{rgb}{0.40,0.80,0.67}
\definecolor{mediumblue}{rgb}{0.00,0.00,0.80}
\definecolor{mediumorchid}{rgb}{0.73,0.33,0.83}
\definecolor{mediumpurple}{rgb}{0.58,0.44,0.86}
\definecolor{mediumsea}{rgb}{0.24,0.70,0.44}
\definecolor{mediumslate}{rgb}{0.48,0.41,0.93}
\definecolor{mediumspring}{rgb}{0.00,0.98,0.60}
\definecolor{mediumturquoise}{rgb}{0.28,0.82,0.80}
\definecolor{mediumviolet}{rgb}{0.78,0.08,0.52}
\definecolor{midnightblue}{rgb}{0.10,0.10,0.44}
\definecolor{mintcream}{rgb}{0.96,1.00,0.98}
\definecolor{mistyrose}{rgb}{1.00,0.89,0.88}
\definecolor{moccasin}{rgb}{1.00,0.89,0.71}
\definecolor{navajowhite}{rgb}{1.00,0.87,0.68}
\definecolor{navyblue}{rgb}{0.00,0.00,0.50}
\definecolor{navy}{rgb}{0.00,0.00,0.50}
\definecolor{oldlace}{rgb}{0.99,0.96,0.90}
\definecolor{olivedrab}{rgb}{0.42,0.56,0.14}
\definecolor{orange1}{rgb}{1.00,0.65,0.00}
\definecolor{orange2}{rgb}{0.93,0.60,0.00}
\definecolor{orange3}{rgb}{0.80,0.52,0.00}
\definecolor{orange4}{rgb}{0.55,0.35,0.00}
\definecolor{orangered}{rgb}{1.00,0.27,0.00}
\definecolor{orange}{rgb}{1.00,0.65,0.00}
\definecolor{orchid1}{rgb}{1.00,0.51,0.98}
\definecolor{orchid2}{rgb}{0.93,0.48,0.91}
\definecolor{orchid3}{rgb}{0.80,0.41,0.79}
\definecolor{orchid4}{rgb}{0.55,0.28,0.54}
\definecolor{orchid}{rgb}{0.85,0.44,0.84}
\definecolor{palegoldenrod}{rgb}{0.93,0.91,0.67}
\definecolor{palegreen}{rgb}{0.60,0.98,0.60}
\definecolor{paleturquoise}{rgb}{0.69,0.93,0.93}
\definecolor{paleviolet}{rgb}{0.86,0.44,0.58}
\definecolor{papayawhip}{rgb}{1.00,0.94,0.84}
\definecolor{peachpuff}{rgb}{1.00,0.85,0.73}
\definecolor{peru}{rgb}{0.80,0.52,0.25}
\definecolor{pink1}{rgb}{1.00,0.71,0.77}
\definecolor{pink2}{rgb}{0.93,0.66,0.72}
\definecolor{pink3}{rgb}{0.80,0.57,0.62}
\definecolor{pink4}{rgb}{0.55,0.39,0.42}
\definecolor{pink}{rgb}{1.00,0.75,0.80}
\definecolor{plum1}{rgb}{1.00,0.73,1.00}
\definecolor{plum2}{rgb}{0.93,0.68,0.93}
\definecolor{plum3}{rgb}{0.80,0.59,0.80}
\definecolor{plum4}{rgb}{0.55,0.40,0.55}
\definecolor{plum}{rgb}{0.87,0.63,0.87}
\definecolor{powderblue}{rgb}{0.69,0.88,0.90}
\definecolor{purple1}{rgb}{0.61,0.19,1.00}
\definecolor{purple2}{rgb}{0.57,0.17,0.93}
\definecolor{purple3}{rgb}{0.49,0.15,0.80}
\definecolor{purple4}{rgb}{0.33,0.10,0.55}
\definecolor{purple}{rgb}{0.63,0.13,0.94}
\definecolor{red1}{rgb}{1.00,0.00,0.00}
\definecolor{red2}{rgb}{0.93,0.00,0.00}
\definecolor{red3}{rgb}{0.80,0.00,0.00}
\definecolor{red4}{rgb}{0.55,0.00,0.00}
\definecolor{red}{rgb}{1.00,0.00,0.00}
\definecolor{rosybrown}{rgb}{0.74,0.56,0.56}
\definecolor{royalblue}{rgb}{0.25,0.41,0.88}
\definecolor{saddlebrown}{rgb}{0.55,0.27,0.07}
\definecolor{salmon1}{rgb}{1.00,0.55,0.41}
\definecolor{salmon2}{rgb}{0.93,0.51,0.38}
\definecolor{salmon3}{rgb}{0.80,0.44,0.33}
\definecolor{salmon4}{rgb}{0.55,0.30,0.22}
\definecolor{salmon}{rgb}{0.98,0.50,0.45}
\definecolor{sandybrown}{rgb}{0.96,0.64,0.38}
\definecolor{seagreen}{rgb}{0.18,0.55,0.34}
\definecolor{seashell1}{rgb}{1.00,0.96,0.93}
\definecolor{seashell2}{rgb}{0.93,0.90,0.87}
\definecolor{seashell3}{rgb}{0.80,0.77,0.75}
\definecolor{seashell4}{rgb}{0.55,0.53,0.51}
\definecolor{seashell}{rgb}{1.00,0.96,0.93}
\definecolor{sienna1}{rgb}{1.00,0.51,0.28}
\definecolor{sienna2}{rgb}{0.93,0.47,0.26}
\definecolor{sienna3}{rgb}{0.80,0.41,0.22}
\definecolor{sienna4}{rgb}{0.55,0.28,0.15}
\definecolor{sienna}{rgb}{0.63,0.32,0.18}
\definecolor{skyblue}{rgb}{0.53,0.81,0.92}
\definecolor{slateblue}{rgb}{0.42,0.35,0.80}
\definecolor{slategray}{rgb}{0.44,0.50,0.56}
\definecolor{slategrey}{rgb}{0.44,0.50,0.56}
\definecolor{snow1}{rgb}{1.00,0.98,0.98}
\definecolor{snow2}{rgb}{0.93,0.91,0.91}
\definecolor{snow3}{rgb}{0.80,0.79,0.79}
\definecolor{snow4}{rgb}{0.55,0.54,0.54}
\definecolor{snow}{rgb}{1.00,0.98,0.98}
\definecolor{springgreen}{rgb}{0.00,1.00,0.50}
\definecolor{steelblue}{rgb}{0.27,0.51,0.71}
\definecolor{tan1}{rgb}{1.00,0.65,0.31}
\definecolor{tan2}{rgb}{0.93,0.60,0.29}
\definecolor{tan3}{rgb}{0.80,0.52,0.25}
\definecolor{tan4}{rgb}{0.55,0.35,0.17}
\definecolor{tan}{rgb}{0.82,0.71,0.55}
\definecolor{thistle1}{rgb}{1.00,0.88,1.00}
\definecolor{thistle2}{rgb}{0.93,0.82,0.93}
\definecolor{thistle3}{rgb}{0.80,0.71,0.80}
\definecolor{thistle4}{rgb}{0.55,0.48,0.55}
\definecolor{thistle}{rgb}{0.85,0.75,0.85}
\definecolor{tomato1}{rgb}{1.00,0.39,0.28}
\definecolor{tomato2}{rgb}{0.93,0.36,0.26}
\definecolor{tomato3}{rgb}{0.80,0.31,0.22}
\definecolor{tomato4}{rgb}{0.55,0.21,0.15}
\definecolor{tomato}{rgb}{1.00,0.39,0.28}
\definecolor{turquoise1}{rgb}{0.00,0.96,1.00}
\definecolor{turquoise2}{rgb}{0.00,0.90,0.93}
\definecolor{turquoise3}{rgb}{0.00,0.77,0.80}
\definecolor{turquoise4}{rgb}{0.00,0.53,0.55}
\definecolor{turquoise}{rgb}{0.25,0.88,0.82}
\definecolor{violetred}{rgb}{0.82,0.13,0.56}
\definecolor{violet}{rgb}{0.93,0.51,0.93}
\definecolor{wheat1}{rgb}{1.00,0.91,0.73}
\definecolor{wheat2}{rgb}{0.93,0.85,0.68}
\definecolor{wheat3}{rgb}{0.80,0.73,0.59}
\definecolor{wheat4}{rgb}{0.55,0.49,0.40}
\definecolor{wheat}{rgb}{0.96,0.87,0.70}
\definecolor{whitesmoke}{rgb}{0.96,0.96,0.96}
\definecolor{white}{rgb}{1.00,1.00,1.00}
\definecolor{yellow1}{rgb}{1.00,1.00,0.00}
\definecolor{yellow2}{rgb}{0.93,0.93,0.00}
\definecolor{yellow3}{rgb}{0.80,0.80,0.00}
\definecolor{yellow4}{rgb}{0.55,0.55,0.00}
\definecolor{yellowgreen}{rgb}{0.60,0.80,0.20}
\definecolor{yellow}{rgb}{1.00,1.00,0.00}

\newcommand{\be}{\begin{equation}}
\newcommand{\ee}{\end{equation}}
\newcommand{\ba}{\begin{eqnarray}}
\newcommand{\ea}{\end{eqnarray}}

\newcommand{\nn}{\nonumber \\}
\newcommand{\x}{\mbox{\boldmath $x$}}

\newcommand{\s}{\mbox{\boldmath $s$}}

\newcommand{\C}{\mbox{\boldmath $C$}}

\newcommand{\D}{\mbox{\boldmath $D$}}
\newcommand{\I}{\mbox{\boldmath $I$}}

\newcommand{\mub}{\mbox{\boldmath $\mu$}}

\newcommand{\k}{\mbox{\boldmath $k$}}

\newcommand{\thetab}{\mbox{\boldmath $\theta$}}

\newcommand{\n}{\mbox{\boldmath $n$}}

\newcommand{\rgl}{\rangle}
\newcommand{\lgl}{\langle}
\newcommand{\de}{\partial}
\newcommand{\Phib}{\mbox{\boldmath $\Phi$}}
\newcommand{\Tr}{{\rm Tr}\,}
\newcommand{\half}{\frac{1}{2}}
\newcommand{\psib}{\mbox{\boldmath $\psi$}}

\newcommand{\calM}{\mbox{${\cal M}$}}

\newcommand{\calL}{\mbox{${\cal L}$}}
\newcommand{\calE}{\mbox{${\cal E}$}}
\newcommand{\calB}{\mbox{${\cal B}$}}
\newcommand{\calS}{\mbox{${\cal S}$}}
\newcommand{\J}{\mbox{\boldmath $J$}}

\def\gs{\mathrel{\raise1.16pt\hbox{$>$}\kern-7.0pt %
\lower3.06pt\hbox{{$\scriptstyle \sim$}}}}         %
\def\ls{\mathrel{\raise1.16pt\hbox{$<$}\kern-7.0pt %
\lower3.06pt\hbox{{$\scriptstyle \sim$}}}}         %

\title[Analytic Likelihood]{
Analytic Methods for Cosmological Likelihoods}
\author[Taylor \& Kitching] {
A. N. Taylor\thanks{ant@roe.ac.uk} \& T. D.
Kitching\thanks{tdk@roe.ac.uk}
 \\
Scottish Universities Physics Alliance (SUPA), Institute for
Astronomy, School of Physics, University of Edinburgh,\\
 Royal
Observatory, Blackford Hill, Edinburgh, EH9 3HJ, U.K.}
\date{}

\pagerange{\pageref{firstpage}--\pageref{lastpage}}

\pubyear{2010}

\begin{document}

\maketitle

\label{firstpage}


\begin{abstract}
We present general, analytic methods for Cosmological Likelihood
analysis and solve the ``many-parameters" problem in Cosmology.
Maxima are found by Newton's Method, while marginalization over
nuisance parameters, and parameter errors and covariances are
estimated by analytic marginalization of an arbitrary likelihood
function with flat or Gaussian priors. We show that information
about remaining parameters is preserved by marginalization.
Marginalizing over all parameters, we find an analytic expression
for the Bayesian evidence for model selection. We apply these
methods to data described by Gaussian likelihoods with parameters
in the mean and covariance. This methods can speed up conventional
likelihood analysis by orders of magnitude when combined with
Monte-Carlo Markov Chain methods, while Bayesian model selection
becomes effectively instantaneous.
\end{abstract}


\section{Introduction}

There is now a Standard Model of Cosmology, $\Lambda$ Cold Dark
Matter ($\Lambda$CDM), which has substantial predictive power but
is highly unsatisfactory from a theoretical viewpoint. The most
serious of these is the unknown nature of the dominant dark energy
component driving the accelerated expansion of the Universe. This
may be due to a new force of nature, or possibly a break-down of
Einstein gravity on large-scales. Without a clear direction of how
to progress beyond a phenomenological picture to a more
fundamental theory, attention is turning to proposing a wide range
of modified or alternative models to the Standard Model and use
observations as a guide to future progress.

To realize this a number of large and challenging observational
programmes are being planned and carried out, for example ESA's
Planck Cosmic Microwave Background mission, the
Canada-France-Hawaii-Telescope Legacy Survey (CFHTLS), ESA's
Visible and Infrared Survey Telescope for Astronomy (VISTA) and
VLT Survey Telescope (VST), the Panoramic Survey Telescope and
Rapid Response System (Pan-STARRS), the Dark Energy Survey (DES),
the Large Synoptic Survey Telescope (LSST), ESA's proposed Euclid
satellite, the NASA/DOE proposed Joint Dark Energy Mission (JDEM),
and the Square-Kilometre Array (SKA). One of the main aims of
these large data-sets is to distinguish between diverse competing
models, some with large parameter-spaces. The $\Lambda$CDM model,
and basic extensions, contains some $18$ parameters, $(\Omega_m,
\Omega_b, \Omega_{\rm de},\Omega_\nu,w_0, w_a, h, A_s, n_s,
\alpha_s, A_T, n_T, \tau,b,f_{\rm NL}, A_{\rm iso},\\\gamma,\eta
)$, covering the dark matter, dark energy, initial conditions and
gravity sectors. Such large parameter-spaces become a problem to
investigate, while fundamental models of dark energy or modified
gravity may have many more parameters which are not well described
by these phenomenological parameters.

The analysis of these large-scale data-sets is not limited by
shot-noise, data volume or the volume of the Universe covered. The
main limitation is our ability to understand and remove, to high
accuracy, systematic effects in the data. For example we may not
precisely know the calibration factor, beam size and shape, or
effect of Galactic foreground contamination in Cosmic Microwave
Background experiments; the calibration and effect of outliers in
photometric redshift surveys; scale-dependent and stochastic bias
in galaxy redshift surveys; calibration of Cosmic Shear or
intrinsic alignment effects in weak lensing surveys; or
environmental effects and evolution in Type Ia supernovae. These
systematic effects are generally parameterized by a set of
nuisance parameters, which themselves must to be constrained by
data. The number of these nuisance parameters can vastly outweigh
the number of cosmological parameters. The size of these large
parameter-spaces for a likelihood analysis is the ``many
parameters" problem.

We also need a systematic approach to discriminating between what
is becoming a large number of competing cosmological models for
dark energy and modified gravity. The Bayesian approach to model
selection is to evaluate the \textit{evidence}, the probability of
model given the data, for all possible cosmological and nuisance
parameter-space. For a large number models, each with a large
number of cosmological and nuisance parameters, this can be an
immense task.

The standard approach to the analysis of cosmological data-sets is
through a likelihood analysis of the model parameter space (e.g.,
Kaiser, 1988; Heavens \& Taylor, 1995; Verde et al., 2003).
Parameter values are given by the maximum, or mean, of the
likelihood function, and parameter errors and covariances are
given by the shape of the marginalized likelihood surface around
the maximum. Since we are not directly interested in nuisance
parameters which characterize  systematic effects, these are
marginalized out. To evaluate the Bayesian evidence we marginalize
over the entire parameter-space, both cosmological and nuisance to
find the probability of the  model.

The likelihood surface can be mapped out numerically using
Monte-Carlo Markov-Chain (MCMC) methods (Gamerman, 1997; MacKay,
2003; Lewis \& Bridle, 2002), where the likelihood distribution is
sampled by a cloud of points whose density follows the likelihood.
Marginalization is then carried out by projecting the points onto
subsets of the parameter-space. As efficient as this is, when the
number of parameters and nuisance parameters becomes large, or
even infinite, this become unfeasible. MCMC is not an efficient or
accurate way to find the maximum of the likelihood, and mean
values are often quoted. The MCMC method can also be sensitive to
the choice of priors, and  insensitive to sharply peaked and
strongly degenerate likelihood surfaces.  Method have evolved to
compensate for this, including using physical parameters (Kosowsky
et al., 2002) or rotating to orthogonal parameter sets (Tegmark et
al., 2004). However, the effect of priors on these spaces is  less
transparent.

An alternative approach to numerical marginalization is to
approximate the likelihood in parameter space as a Gaussian and
analytically marginalize (Bretthorst, 1988; Gull, 1989; Bridle et
al., 2002; MacKay, 2003). Bridle et al. (2002)  apply this method
in cosmology to marginalize over nuisance parameters appearing in
the mean of a Gaussian likelihood. This approach is exact when the
parameters are Gaussian distributed such as the amplitude of the
mean, and this is publicly available in
CosmoMC\footnote{http://cosmologist.info} (Lewis \& Bridle, 2002).
An analytic marginalization method has also been developed for
evaluating the Bayesian evidence, using the saddle-point, or
Laplace, approximation to marginalize over all parameters around
the peak of the likelihood (e.g., MacKay, 2003; Trotta, 2008).
However, this does not evaluate the absolute evidence. There is no
general treatment of analytic marginalization which will
accommodate both of these, and even more general, situations. In
this paper we present a new, self-consistent and general framework
in which to maximize and marginalize over an arbitrary likelihood
function, to remove nuisance parameters, estimate marginalized
projections of parameter-space, and derive an analytic expression
for the Bayesian evidence.

The paper is set out as follows. In Section 2 we describe
Likelihood methods for parameter estimation and set out the
general approach for maximization and marginalization over
nuisance parameters for an arbitrary likelihood function with flat
or Gaussian priors. We show that the marginalized likelihood
function preserves information on cosmological parameters. In
Section 3 we show how to apply the method to the specific case of
a multivariate Gaussian-distributed data where the cosmological
and systematic information is contained in the mean and
covariance. In Section 4 we present some applications:
marginalization over an amplitude, projections of parameter-space,
and semi-analytic marginalization. We show how our methods can
applied to find a solution to the problem of Bayesian evidence in
Section 5, and discuss some aspects of model selection in
model-space. Finally, in Section 6 we present our conclusions.

\section{Analytic Likelihood Analysis}

Assuming a model, $\calM$, for a cosmological dataset, $\D$, which
is parameterized by a set of $N_p$ parameters, $\thetab$, the
conditional probability distribution of the data is given by the
likelihood function, $L=p(\D|\thetab,\calM)$. We can transform
from the likelihood function to the posteriori probability for the
parameters given the data, $p(\thetab|\D,\calM)$, using Bayes'
Theorem;
 \be
    p(\thetab|\D,\calM) = \frac{L(\D|\thetab,\calM) p(\thetab|\calM)}{
  p(\D|\calM)},
  \label{bayes_like}
 \ee
where $p(\thetab|\calM)$ is the prior distribution of the
 parameters assumed before the analysis. The
normalizing distribution, $p(\D|\calM)$, is called the {\em
evidence}. Priors are commonly assumed to be either flat, where
the distribution is a top-hat with constant value over some
parameter range and zero outside, or Gaussian with a mean
constrained by earlier experiments. The posterior distribution is
then maximized with respect to the $N_p$ cosmological parameters
in the model. Marginalization of the posteriori or likelihood
function is required if we have a subset of $M$ parameters,
$\psib$, which we want to integrate over;
 \be
        p(\thetab|\calM) = \int \! d^{M} \! \!\psi\, p(\thetab,\psib|\calM).
 \ee
The $\psib$-parameters may be nuisance parameters which
characterize some systematic effect, or some of the cosmological
parameters, $\thetab$, whose effect we want to integrate over when
we do not have an accurate understanding of the effect (for
example the normalization of galaxy perturbations due to galaxy
bias). We may also want to project out the likelihood surface to
lower dimensions to study the distribution, or even marginalize
over all of the $N_p+M$ nuisance and cosmological parameters if we
want to estimate the evidence.

Now consider an arbitrary likelihood function,
$L(\D|\Phib,\calM)$, which depends on a set of cosmological
parameters, $\thetab$, and on a set of marginalization parameters,
$\psib$, which we want to integrate over, where we have combined
all parameters into $\Phib=(\thetab,\psib)$. We begin by defining
the log-likelihood, $\calL$, of the likelihood function
 \be
    \calL = -2 \ln L.
 \ee
This can be expanded around an arbitrary point, $\Phib_0$, in the
full parameter-space to second-order
 \be
 \label{expand}
        \calL   = \calL_0 + \delta \Phi_\mu  \calL_\mu
                +\half \delta \Phi_\mu \delta
                    \Phi_\nu \calL_{\mu\nu} ,
 \ee
where $\calL_\nu = \de_\nu \calL_0$ and $\calL_{\nu\mu}=\de_\nu
\de_\mu \calL_0$ are evaluated at $\Phib_0$, and where we denote
derivatives with respect to a nuisance parameter by Greek indices.

\subsection{Maximizing the likelihood}

We first want to find the minimum of the log-likelihood function
in the full $N_P+M$ cosmological and nuisance parameter-space.
Differentiating equation (\ref{expand}) with respect to the
parameters and setting the gradient to zero, we find the
displacement between the fiducial point and the peak of the
likelihood is
 \be
   \delta \Phi_\mu = - \calL_\nu \calL_{\nu \mu}^{-1}.
 \ee
If the likelihood is close to Gaussian we can find the maximum of
the likelihood in a single step. If the likelihood is
non-Gaussian, but smooth, we can iterate towards the peak. This is
Newton's method for finding the peak of the likelihood (e.g.,
Press et al., 1989).

\subsection{Analytic Marginalization}
\label{AM}

We now want to marginalize over the $\psib$ nuisance parameters.
Expanding the likelihood in the $\psib$-parameters yields;
 \be
 \label{eq4}
        \calL   = \calL_0 + \delta \psi_\alpha  \calL_\alpha
                +\half \delta \psi_\alpha \delta
                    \psi_\beta \calL_{\alpha\beta} ,
 \ee
where the indices $\alpha$ and $\beta$ refer to nuisance
parameters. Analytically marginalizing over $\psib$ (see Appendix
A for details), assuming a non-zero flat prior in the volume
$V_{\psi}$ of $\psib$-space, $p(\psib|\calM)=1/V_{\psi}$, yields
 \be
        \calL = \calL_0
            - \half \calL_\alpha \calL_{\alpha \beta}^{-1} \calL_\beta
            + \Tr \ln \left( V^{2/M}_\psi \calL_{\alpha\beta}
            \right) ,
            \label{maglike}
 \ee
where we have dropped an unimportant constant of $- M \ln ( 4 \pi
)$. This is the marginalized log-likelihood function. In the first
term, $\calL_0=\calL(\thetab|\psib=\psib_0)$ is the conditional
likelihood at fixed $\psib$.

The second term in equation (\ref{maglike}), which is quadratic in
$\calL_\alpha$, has an intuitive meaning. Although we have fixed
the values of $\psib=\psib_0$ at their maximum in the full
parameter space, and where the gradient is zero, the likelihood is
still a function of the remaining parameters, $\thetab$. As we
move in parameter space away from the maximum along one of the
directions of $\thetab$, the peak will move away from $\psib_0$,
unless the parameters are uncorrelated, and the gradient
$\calL_\alpha$ will be non-zero. This term then describes the full
shape of the likelihood and the coupling between the marginalized
parameters and the remaining parameters. Its presence removes the
dependence of the likelihood on the marginalized parameters, and
widens the distribution.

The third, well-known, term accounts for the volume of
marginalized parameter-space with significant likelihood, and is
called the \textit{Occam factor.} The presence of the curvature of
the log-likelihood, through $\calL_{\alpha\beta}$, shows that this
expression is sensitive to information in the data itself about
the systematic nuisance parameters. Note that we have made no
assumptions about the form of the likelihood function in
$\thetab$-space, only that we can approximate the peak of the
likelihood function in the marginalized $\psib$-parameter space by
a multivariate Gaussian. Analytic marginalization does not suffer
from prior boundary problems, since the full likelihood space is
marginalized over, and infinitely resolves the peak of the
likelihood.

We can derive the marginalized likelihood in a second, more
illuminating, way. We can use the expansion given by equation
(\ref{eq4}) to find the displacement of a fixed point in nuisance
parameter-space from the peak of the likelihood,
 \be
    \delta \psi_\alpha = - \calL_\beta \calL_{\alpha \beta}^{-1}.
  \ee
Substituting this back into equation (\ref{eq4}) we find that
maximum value of the likelihood is
 \be
  \calL_{\rm max} = \calL_0
            - \half \calL_\alpha \calL_{\alpha \beta}^{-1}
            \calL_\beta.
            \label{lmax}
 \ee
The first two terms in equation (\ref{maglike}) are just the
maximum likelihood value, while the third term is just the width
of the likelihood curve. This shows us that the marginalized
likelihood is independent of the choice of $\psib_0$, when
$\calL(\psib)$ is Gaussian, since the second term in equation
(\ref{lmax}) corrects the likelihood estimated at $\psib_0$ to the
value at the peak. In Appendix B we derive the mean and variance
of the likelihood from its Generating Function.

Analytic marginalization  preserves information about cosmological
parameters. Expanding equation (\ref{maglike}) to lowest order in
the remaining cosmological parameters, $\Delta \thetab$, around
the peak of the ensemble averaged likelihood keeping the curvature
$\calL_{\alpha\beta}$ fixed at its expectation value, we find
 \be
    \calL = \calL_0 +  \Delta \theta_i \Delta \theta_j
\left[\lgl \calL_{ij}\rgl -   \lgl \calL_{i\alpha} \rgl
  \lgl \calL_{\alpha\beta}\rgl^{-1}  \lgl \calL_{\beta
  j}\rgl\right],
 \ee
where Arabic indices $i$ and $j$ indicate cosmological parameters.
Here we can identify the Schur complement (e.g., Zhang, 2005) of
the marginalized Fisher information matrix for cosmological
parameters,
 \be
        F_{ij}^M = F_{ij}- F_{i\alpha} F^{-1}_{\alpha
        \beta} F_{\beta j}, \label{schur}
 \ee
where
 \be
 F_{\mu \nu} = \half  \lgl \calL_{\mu\nu} \rgl
 \ee
is the full $N_p+M$-dimensional Fisher matrix (see, e.g., Tegmark,
Taylor \& Heavens, 1997) for cosmological parameters and
systematic nuisance parameters. The indices $(\mu,\nu)$ extend
over all $(i,j)$ and $(\alpha,\beta)$. Equation (\ref{schur}) is
identical to the Fisher matrix found by maximizing the
pre-marginalized likelihood and then marginalizing over the
nuisance parameters. Hence, at the level of Fisher Matrices, no
information is lost by analytic marginalization.

When we have a Gaussian prior on the nuisance parameters the
log-likelihood becomes
 \be
    \calL = \calL_0 + \delta \psi_\alpha  \calL_\alpha +\half \delta
    \psi_\alpha [\calL_{\alpha\beta}+2C_{\alpha\beta}^{-1}] \delta
    \psi_\beta + \Tr \ln C_{\alpha\beta},
 \ee
where $C_{\alpha\beta}$ is the prior covariance matrix. The
maximum is now found at
 \be
  \delta \psi_\alpha = - \calL_\beta [\calL_{\alpha \beta}
        + 2C_{\alpha \beta}^{-1}]^{-1},
  \ee
while marginalization leads to
 \be
        \calL = \calL_0 - \half\calL_\alpha [
        \calL_{\alpha \beta}+ 2 C^{-1}_{\alpha\beta}]^{-1}
        \calL_\beta + \Tr \ln
        \left( \delta^K_{\alpha\beta}
        + \half C_{\alpha \delta}\calL_{\delta\beta}\right).
 \ee

\section{Gaussian Likelihoods}

Let us assume the statistical properties of the data, $\D$, can be
modelled by a multivariate Gaussian distribution,
$L(\D|\thetab,\psib)$ which depends only on a mean,
$\mub(\thetab,\psib)=\lgl \D \rgl$, and a covariance matrix,
$\C(\thetab,\psib) = \lgl \Delta \! \D \Delta\! \D^t \rgl$, where
$ \Delta \!\D = \D - \mub(\thetab,\psib) $ is the variation of the
data about the mean. By definition $ \lgl \Delta \! \D \rgl =0. $
The Gaussian log-likelihood function is given by
 \be
    \calL_0=\Delta \D \C^{-1} \Delta \D^t + \Tr \ln \C.
 \ee
The cosmological and nuisance parameters can appear in both the
mean of the data values, or in the covariance. We consider each in
turn, starting with parameters in the mean.

\subsection{Parameters in  the mean}

If the nuisance parameters are in the mean, $\mub=\mub(\psib)$,
and we assume a flat prior on marginalization parameters, the
gradient and curvature of the log-likelihood in parameter-space is
\ba \label{pinmean}
    \calL_\alpha &=& - 2\Delta \D^t \C^{-1} \mub_\alpha , \\
    \calL_{\alpha\beta} &=& 2\left(\mub_\alpha \C^{-1}\mub^t_\beta
    - \Delta \D^t \C^{-1} \mub_{\alpha\beta}\right) .
\ea
The expectation value of the slope is $\lgl \calL_\alpha \rgl =0$,
while the expectation value  of the curvature around the peak in
parameter-space is,
 \be
        \lgl \calL_{\alpha\beta}\rgl = 2 F_{\alpha \beta} =
        2 \mub_\alpha \C^{-1} \mub^t_\beta.
 \ee
If we choose to use the Fisher Matrix for the local curvature, the
maximum of the Gaussian likelihood function lies at
 \be \label{peak}
    \Phi^{\rm max}_\nu=\Phi^0_\nu +F^{-1}_{\mu\nu}\Delta \D^t \C^{-1}
    \mub_{\mu}
 \ee where
where $\Phib^0$ is an arbitrary point in parameter-space. Since
the  curvature is approximated by the Fisher matrix, this is a
quasi-Newtonian method. Again if the likelihood is Gaussian in
parameter-space, this is exact, and if not some iteration is
required.

Marginalizing over the nuisance parameters assuming a flat prior,
we find the likelihood function is again a Gaussian,
 \be
        \calL =
                \Delta \D \C_M^{-1} \Delta \D^t
                + \Tr \ln V^{2/M}_{\psi} F_{\alpha \beta},
                \label{marg_gauss}
 \ee
where the marginalized data covariance matrix, $\C_M$ is given by
 \be
        \C_M = \lgl \Delta \D \Delta \D^t \rgl_M =
            \left(\C^{-1}- \C^{-1}\mub^t_\alpha
            F_{\alpha \beta}^{-1}
            \mub_\beta\C^{-1}\right)^{-1}.
 \ee
If we assume the curvature is given by its expectation value, the
constant term, $\ln \det V^2_\psi F_{\alpha \beta}$ in equation
(\ref{marg_gauss}), can be dropped and we can identify $\calL$
with the  $\chi^2$-statistic and all our results still hold. Note
that in these expressions the parameter-dependence only appears in
the mean in $\Delta \D= \D - \mub(\thetab, \psib_0)$. Everything
else is fixed at the fiducial values, $\thetab_0$ and $\psib_0$.
We can also see from this solution that there is a requirement on
the marginalized covariance matrix that it is positive definite, $
\Delta \D \C_M^{-1} \Delta \D^t > 0, $ in order that the
likelihood function has a maximum bound, however this is always
true.


If we assume a Gaussian prior on the nuisance parameters, the
marginalized data covariance matrix is regularized and can be
simplified using the Woodbury matrix identity (Woodbury, 1950) so
that
 \ba
        \C_M &=&
        \left(\C^{-1}- \C^{-1} \mub^t_\alpha
            [F_{\alpha \beta} + C_{\alpha\beta}^{-1}]^{-1}
            \mub_\beta \C^{-1}\right)^{-1}
            \label{marg_full}
     \\
            &=& \C + C_{\alpha \beta} \mub_\alpha \mub^t_\beta,
     \label{marg_wood}
 \ea
where the last expression is explicitly positive-definite.
Equations (\ref{marg_full}) and (\ref{marg_wood}) have previously
been derived by Bridle et al. (2002) using a somewhat different
method for marginalizing over a Gaussian likelihood with a
Gaussian prior and nuisance parameters in the mean. If we include
a prior on nuisance parameters the log-likelihood function becomes
 \be
    \calL = \Delta \D \C_M^{-1} \Delta \D^t + \Tr \ln \C_M,
 \ee
again up to an unimportant normalization constant. We note that
even if the cosmological parameters do not affect the covariance,
the marginalized covariance, $\C_M$, will gain a dependence on
cosmological parameters through the mean.

\subsection{Parameters in the covariance}

If the parameters are in the data covariance matrix,
$\C=\C(\thetab,\psib)$, the derivatives of the log-likelihood are
 \ba
    \calL_\alpha &=& - \Tr \left( \de_\alpha \!\ln \C \,
                        \Delta \!\ln \C \right),
    \label{covgrad} \\
    \calL_{\alpha\beta} &=&
                \Tr \big[(\de_\alpha \!\ln \C)(\de_\beta  \ln \C)
                (\I+2\Delta \! \ln \C ) \nn
                & &
          -\C^{-1}(\de_\alpha \de_\beta \C)  \Delta \! \ln \C\big].
 \ea
where $\de_\alpha \! \ln \C = \C^{-1} \de_\alpha \C $, $\Delta \!
\ln \C =  \Delta \D  \C^{-1} \Delta \D^t - \I$ and $\lgl \Delta \!
\ln \C \rgl = 0$. The expectation values of the gradient is $\lgl
\calL_\alpha \rgl = 0$ while the expectation of the curvature is
given by,
 \be
        \lgl \calL_{\alpha\beta}\rgl = 2 F_{\alpha\beta}=
        \Tr[(\de_\alpha \! \ln \C)(\de_\beta \ln \C)].
 \ee
If we assume the curvature is given by its expectation value we
find the peak is at
 \be
    \delta \Phi_\nu = \half F^{-1}_{\nu\mu}\,
    \Tr \left( \de_\mu \!\ln \C \, \Delta \! \ln \C \right),
 \ee
from the fiducial point in $\Phib$-space. For a single-step
estimate of the peak, this is equivalent to Tegmark's (1997)
Quadratic Estimator. The analytically marginalized log-likelihood
is
 \be
        \calL = \calL_0 - \frac{1}{4} \calL_\alpha F_{\alpha\beta}^{-1}
        \calL_\beta + \Tr \ln  V^{2/M}_\psi F_{\alpha \beta},
 \ee
where $\calL_\alpha$ is given by equation (\ref{covgrad}). To
change the prior to a Gaussian we again make the substitution
 \be
        \calL = \calL_0 - \frac{1}{4}\calL_\alpha [F_{\alpha\beta}+
        C_{\alpha \beta}^{-1}]^{-1}
        \calL_\beta + \Tr \ln \left( \delta^K_{\alpha \beta}
        + C_{\alpha \beta} F_{\alpha \beta} \right),
 \ee
Again, we require that $\calL >0$ to bound the likelihood
function.

\section{Applications}

Having calculated the marginalized likelihoods for
Gaussian-distributed data with parameters in both mean and
covariance matrix, we now turn to two examples: marginalization
over nuisance parameters and projections of the likelihood
function in parameter-space.

\subsection{Systematic Nuisance Parameters}

A simple, and well-known, example of a nuisance parameter is the
normalization of the mean with a flat prior. This is an
interesting case since the analysis is exact. Let the mean be
given by $\mub = A \mub_0$, where the Fisher matrix for the
amplitude, $A$, found from the data is given by $F_{AA}=(1/A^2)
\Tr [\mub \C^{-1} \mub^t]$, then
 \be
        \C_M = \left(\C^{-1} - \frac{\C^{-1} \mub^t
        \mub \C^{-1}}{\Tr[\mub \C^{-1} \mub^t]}\right)^{-1}
 \ee
and the peak is found from equation (\ref{peak}). If we assume the
covariance is diagonal, $C_{ij}=\sigma_i^2 \delta^K_{ij}$, then
the
log-likelihood becomes
 \be
    \calL = \sum_i \frac{\Delta D_i^2}{\sigma_i^2}
    -\left( \frac{1}{
     \sum_k \mu_k^2/\sigma^2_k}\right)
    \left( \sum_i \frac{\Delta D_i \mu_i}{\sigma_i^2}\right)^2.
 \ee
If we assume further that the mean values are Gaussian-distributed
power spectra, $\mu_k = P_k$, their variance is given by
$\sigma^2_{k} = 2 P_k^2 $, and the log-likelihood is
 \be
  \calL =
        \half \sum_{\k} \left( \Delta \ln P_k -
    \overline{ \Delta \ln P_k}\,\right)^2.
 \label{cal_pow}
 \ee
In the last expression $\Delta \ln P_k = [\widehat{P}_k-
P_k(\thetab)]/P_k$, where $\widehat{P}_k$ is the measured power,
$\overline{x} = (1/N_{\rm D}) \sum_k x_k$ and $N_{\rm D}$ is the
number of data points. Hence the log-likelihood is
positive-definite, and minimizing $\calL$ is equivalent to
minimizing the variance of $\Delta \! \ln P_k$. This expression
makes sense as the second term removes any dependence on the best
estimate of the calibration off-set from the likelihood. Equation
(\ref{cal_pow}) has an immediate cosmological application for
removing the dependence of a linear galaxy bias on parameters
estimated from the galaxy power spectrum, assuming the power
spectrum pass-bands are independent.

More generally we find the marginalized likelihood for multiple
parameters is given by
 \be
        \calL = \half \left[\sum_{\k} |\Delta \!\ln P_k|^2 -
            \half \calL_\alpha F_{\alpha \beta}^{-1}
            \calL_\beta \right],
    \label{like_mean}
 \ee
where the Fisher matrix and gradient of the log-likelihood are
 \ba
        F_{\alpha\beta} &=& \half \sum_{\k}
        (\de_\alpha \! \ln P_k) ( \de_\beta \ln P_k), \\
        \calL_\alpha &=&-\sum_{\k} \Delta
            \!\ln P_k \,\de_\alpha \ln P_k,
 \ea
and the peak of the likelihood is at
 \be
    \delta \Phi_\mu = - \half F_{\mu \nu}^{-1} \calL_\nu.
 \ee
 If we want to include noise in these expressions, we can
do so by substituting $P_k \rightarrow P_k + N(r)$, where $N(r)$
is the noise power, which may depend on position within the
survey. For example in galaxy redshift surveys, $N(r)=1/\bar
n(r)$, and we should extend the summation over $k$ to $\Tr
\rightarrow \sum_{k} \int d^3r $. In the continuum limit we would
substitute $\Tr= \int d^3 k/(2 \pi)^3$ (see, for example, Taylor
\& Watts, 2001). For CMB or weak lensing analysis on the sky, we
should substitute $P_k \rightarrow C_\ell$ and $\Tr \rightarrow
\sum_\ell (2 \ell+1)$, where we have implicitly assumed
statistical isotropy and summed over the $2 \ell+1$ azimuthal
modes. Finally, for 3-D Cosmic Shear (e.g., Heavens, Kitching \&
Taylor, 2006), where the covariance matrix is $\C =
C^{\gamma\gamma}_\ell(z,z')$ we  substitute $\Tr \rightarrow
\sum_\ell (2\ell+1) \int dz dz'$.

If the parameter appear in the covariance matrix, and the data has
a Gaussian distribution, the log-likelihood distribution is given
by
 \be
    \calL_0 = \Tr \left(\widehat{\C} \C^{-1} + \ln \C \right)
    = \Tr (\Delta \!\ln \C + \ln \C) + N_D,
 \ee
where $\widehat{\C}= \Delta \D \Delta \D^t$, and $N_D$ is the
number of data-points used. If again we use the example of
marginalization over the normalization of the covariance matrix,
$\C = A \C_0$, where the Fisher matrix is $F_{AA}=N_D/2A^2$, the
marginalized likelihood is
 \be
    \calL =
    \Tr (\Delta \!\ln \C + \ln \C)
    -\frac{1}{N_D} \Tr \left[ \Delta \!\ln \C \Delta \!\ln \C\right] + N_D.
 \ee
For a diagonal covariance matrix the marginalized log-likelihood
with parameters in the covariance can be written
 \be
        \calL = \sum_{\k} \left(\frac{ \widehat{P}_k}{P_k}
        + \ln P_k \right) -
        \frac{1}{4}\calL_\alpha F_{\alpha \beta}^{-1} \calL_\beta .
 \ee
Despite the different form of the term $\calL_\alpha
\calL_{\alpha\beta}^{-1}\calL_\beta$ when the parameters appear in
the data covariance matrix, in this limit this term is the same as
when the parameter appear only in the mean (c.f. equation
\ref{like_mean}).


\begin{figure}
 \includegraphics[width=\columnwidth]{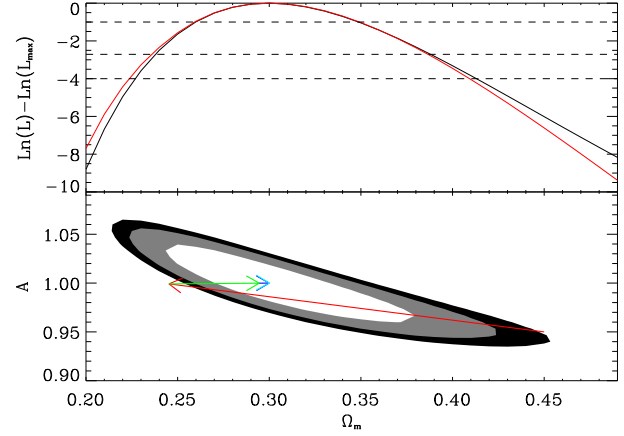}
 \caption{\em Example of marginalization over a nuisance parameter.
   The lower panel shows the two-parameter $1$- $(68.3\%)$,
   $2$- $(90\%)$ and $3$-$\sigma$ ($99.9\%$) contours in white, gray
   and black for the matter-density parameter, $\Omega_m$, and
   a nuisance power-spectrum normalization parameter, $A=b\sigma_8$,
   for a measurement of the matter
   power spectrum for a survey covering an effective volume of
   $19.7h^{-3}{\rm Gpc}^3$ with negligible shot-noise. The solid
   line show the convergence to the maximum likelihood.
   The upper panel compares the one-parameter marginalized $\Omega_m$
   constraint for full numerical marginalization (black) with
   analytic marginalization using equation (\ref{cal_pow}) (red), the
   difference between these lines, even in this non-Gaussian case, is
   small. The dashed lines show the one-parameter
   $1$-, $2$- and $3$-$\sigma$ limits (assuming a Gaussian likelihood).}
 \label{pk}
\end{figure}

\begin{figure*}
  \includegraphics[width=18.cm]{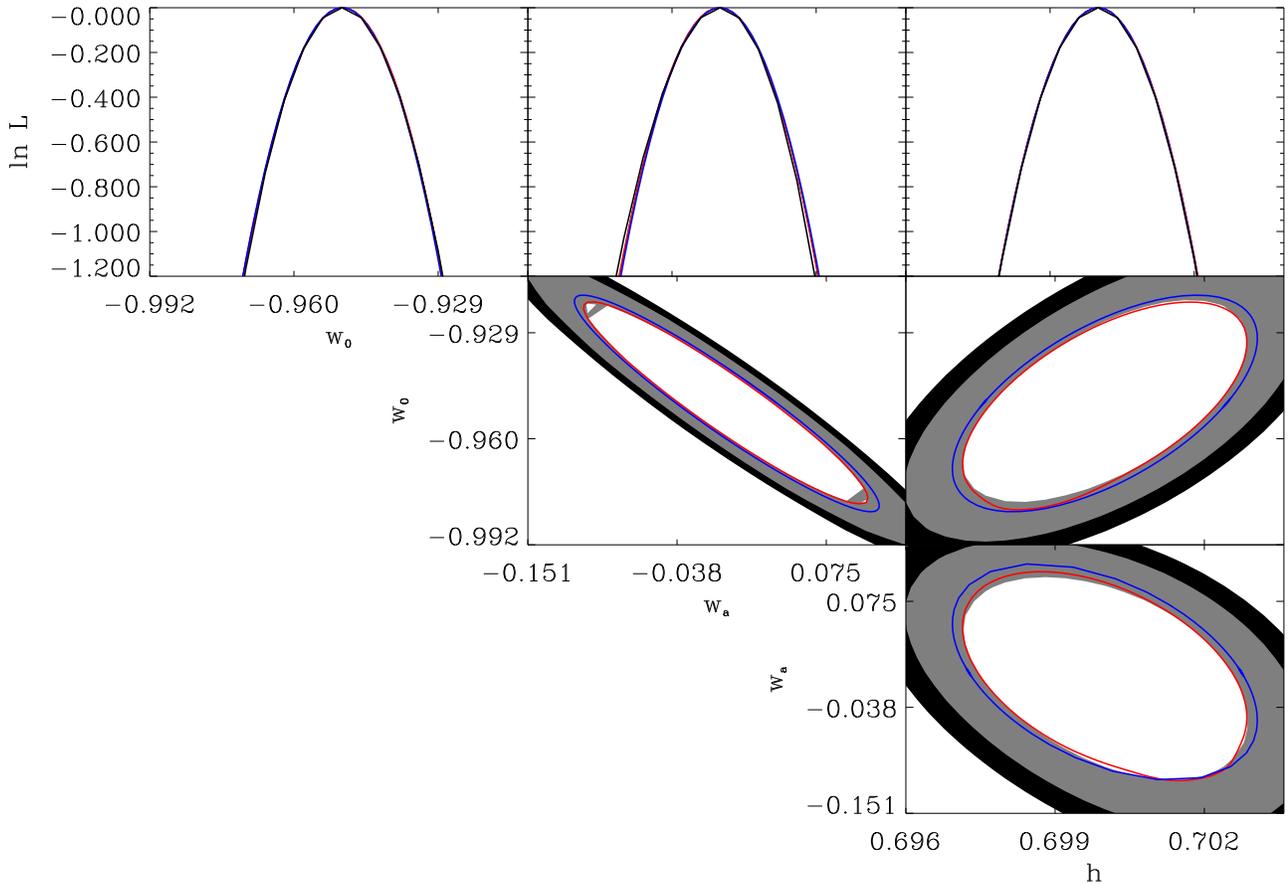}
  \caption{\em Projected cosmological $3$-parameter space for a
    Euclid-type (20,000 square degrees, median redshift of $z=0.8$)
   gravitational lensing survey. Grey contours are 1- 2-
   and 3-$\sigma$ levels using analytic marginalization over the extra
   parameters,
   solid blue lined ellipses are the 1-$\sigma$ contours using the
   Fisher matrix approximation to the projected
   likelihood surface, solid red ellipses are the 1-$\sigma$ fully
   marginalized constraints. The upper panels show the 1D
   marginalized
   likelihoods for the analytic marginalization (black), the Fisher
   approximation (blue) and for a full numerical marginalization (red).}
 \label{like}
\end{figure*}

\subsubsection{Galaxy clustering}
\label{powerspec}

In Figure \ref{pk} we show the likelihood, $L(\Omega_m,A)$, for a
joint measurement of the matter-density parameter, $\Omega_m$, and
galaxy clustering amplitude, $A = b \sigma_8$, from the galaxy
power spectrum, $P_g(k)$. Here $b$ is a linear bias parameter and
$\sigma_8$ the variance of matter clustering in spheres of $8
h^{-1}{\rm Mpc}$.  The matter power spectrum is generated using
the Eisenstein \& Hu (1997) parameterization with a Smith et al.
(2003) non-linear correction, and we have ignored the effect of
redshift-space distortions.. We have assumed a fixed Hubble
parameter, hence $\Omega_m$ determines the linear break-scale in
the matter power-spectrum, and amplitude of nonlinear corrections.
We assume a fiducial model with $\Omega_m=0.3$ and $b\sigma_8=1$.
The error on the measured power is assumed to be sample-dominated,
with negligible shot-noise, given by $\sigma(k)=2 \pi P(k)
/\sqrt{V k^3 d \ln k}$ (e.g., Tegmark 1997), where we have assumed
$V=19.7 h^{-3}{\rm Gpc}^3$ and spectroscopic redshifts and no
redshift-space distortion. We include a wavenumber range up to
$k_{\rm max}=100 h$Mpc$^{-1}$. We show in the lower 2-parameter
distribution how Newton's Method convergence to the maximum
likelihood. It is clear that after approximately $3$--$4$
iterations the maximum likelihood is covered, even in this case of
a highly non-Gaussian likelihood surface.

Since the galaxy bias parameter is poorly known, it is useful to
marginalize over the amplitude when estimating $\Omega_m$. The
upper plot in Figure \ref{pk} shows the projected 1-d marginalized
likelihood for $\Omega_m$, for both numerical marginalization over
the amplitude (black line), and using the analytic marginalization
result given by equation (\ref{cal_pow}) (red line). The analytic
result accurately reproduces the full numerical result for the
1-, 2- and 3-$\sigma$ errors, even though there is
some non-Gaussianity in the $\Omega_m$--$A$ plane.

\subsection{Projection of parameter-space}

Another application for analytic marginalization is in the
projection of parameter-space. Usually the maximum likelihood
parameter values are quoted along with the marginalized errors and
marginalized parameter covariances.  Sometimes the mean of a
parameter, marginalized over all other parameters, is also quoted
(e.g., Spergel et al., 2003), and the 2-D projected
parameter-space plotted to illustrate non-Gaussianity. We can
again use analytic marginalization to do this for us.


\subsubsection{Dark energy parameters from 3-D Cosmic Shear}

In Figure \ref{like} we show the predicted projected likelihood
space estimated on a grid for a set of $3$ cosmological
parameters, $(w_0, w_a, h)$ where $w(a)=w_0+(1-a)w_a$ is the dark
energy equation of state, $p=w(a) \rho$, and $h=H_0/100 {\rm
km}s^{-1}{\rm Mpc}^{-1}$ is the reduced Hubble parameter. The
fiducial maximum-likelihood values are $w_0=-0.95$, $w_a= 0$, and
$h= 0.7$, and we have assumed a 3-D tomographic cosmic shear
analysis with the proposed Euclid satellite mission (Refregier et
al., 2006), covering $20$,$000$ square degrees with median
redshift $z=0.8$ and a number density of $35$ galaxies per sqaure
arcminute. 
The upper row in Figure \ref{like} compares the
analytically marginalized 1-D parameter distribution with
numerical marginalization over the remaining 2-D likelihood
surface and the Fisher matrix prediction. We see that analytic
marginalization is indistinguishable from numerical
marginalization. The lower panels show the projected 2-D
likelihood surface for analytic marginalization (solid
white/grey/black 1-, 2-, 3-$\sigma$ regions) along with the
two-parameter $1$-$\sigma$ (68.3\%) likelihood contours estimated
from the Fisher matrix approximation (blue ellipse), and a contour
for the numerical marginalization (red ellipse). It can be seen in
all panels that the analytic marginalized likelihood surface is in
excellent agreement with the numerical marginalization,
reproducing even small departures from the Fisher Matrix
approximation. While results will clearly depend on which
parameters are in the likelihood analysis, this does suggest that
for large numbers of parameters, the marginalization will tend
towards a Gaussian distribution, since any departures from
Gaussianity will be averaged out.

In Figure \ref{multi_like} we extend the comparison to an
8-parameter cosmological model. In this example the qualitative
differences between the analytic marginalization result and are
clear. In some 2-D parameter spaces for example ($\Omega_b$,$h$)
there is significant non-Gaussianity, however in others such as
($w_0$,$w_a$) the 2-D parameter space is very Gaussian. In such
circumstances analytic marginalization could be used to
marginalize over Gaussian parameter combinations and a numerical
marginalization used to capture any non-Gaussian behaviour.

\subsection{Semi-analytic marginalization}

\begin{figure*}
  \includegraphics[width=2.\columnwidth]{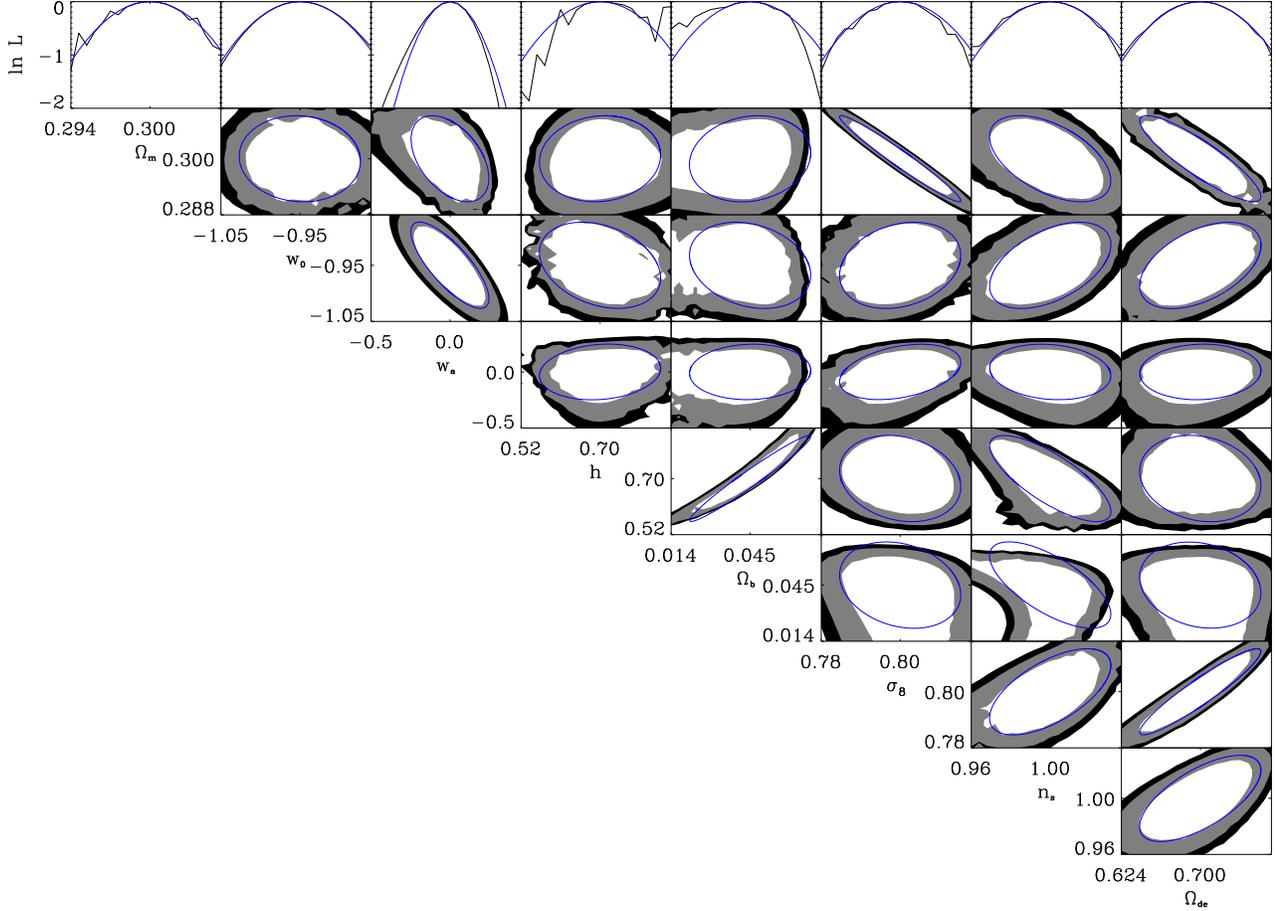}
  \caption{\em Projected cosmological $8$-parameter space for a
    Euclid-type (20,000 square degrees, median redshift of $z=0.8$)
   gravitational lensing survey. The upper panel show the 1D parameter
   constraints using analytic marginalization (black) and the Fisher
   matrix approximation (blue, dark gray). The other panels show the
   2D parameter constraints. Grey contours are 1- 2-
   and 3-$\sigma$ levels using analytic marginalization over the extra
   parameters,
   solid blue ellipses are the 1-$\sigma$ contours using the
   Fisher-matrix approximation to the projected
   likelihood surface, solid red ellipses are the 1-$\sigma$ fully
   marginalized. }
 \label{multi_like}
\end{figure*}

Non-Gaussianity is significant for some parameters and so we
propose an algorithm for \emph{semi-analytic marginalization}.
Having found the $N_p+M$-parameter maximum-likelihood peak by a
quasi-Newton solution,
 \be
    \delta \Phi_\nu = - \half F^{-1}_{\mu\nu} \calL_\mu,
 \ee
we can use MCMC to plot out the 1- and 2-D parameter likelihood
distributions, analytically marginalized over all other
parameters. The non-Gaussian parameters can be removed from the
analytic marginalization and numerically marginalized over with
MCMC. If new, non-Gaussian parameters appear we can numerically
marginalize over them until stability is reached. This process may
end up running MCMC on all parameters -- but in many cases some,
if not many, of the parameters will be close to
Gaussian-distributed in parameter-space with just a few
non-Gaussian parameters needing numerical marginalization. In this
case the time spent mapping parameter space can be decreased
significantly. We assume the time to run a full MCMC analysis in a
$N_p$-parameter space is
 \be
    T_{\rm MC} = \Delta t_{\rm  MC} N_p \ln N_p,
 \ee
where $\Delta t_{\rm  MC}$ is the time to run one point in the
MCMC chain. If $M$ of these parameters can be analytically
marginalized over, a semi-analytic marginalization scheme will
take
 \be
    T_{\rm SAM} = \Delta t_{\rm MC} (N_p-M) \ln (N_p-M)
        + \Delta t_{\rm F} M,
 \ee
where $\Delta t_{\rm F} \ll \Delta t_{\rm MCMC}$ is the time taken
to estimate the Fisher matrix. Clearly if all parameters are well
approximated by a multivariate Gaussian, the main effort is in
finding the peak of the likelihood, since we already know the
Fisher matrix. For example in our 8-parameter cosmological model
(Figure \ref{multi_like}), only the baryon density, $\Omega_b$,
and the scalar spectral index, $n_s$, show significant deviations
from Gaussianity. This implies we can reduce the computation time
by a factor of $12$. If we have a model with an additional 200
nuisance parameters, all of which can all be marginalized over,
this is a reduction of around 67. Even if MCMC has be to
extensively used to map out the parameter-space, analytic
marginalization can also be used to map the MCMC proposal
distributions more accurately than a Fisher Matrix approximation.

\section{Model selection and the Bayesian Evidence}

\subsection{The Bayesian Evidence}

Having explored analytic methods for maximizing and marginalizing
in a likelihood analysis, we now turn to the problem of model
selection. For model selection we need to find the probability of
the most likely model given the data, $p(\calM|\D)$. From Bayes'
Theorem we find (see e.g., Liddle 2009, Trotta 2008)
 \be
    p(\calM|\D) = \frac{p(\D|\calM) p (\calM)}{p(\D)},
    \label{model_bayes}
 \ee
where the probability $p(\D|\calM)$ can be identified as the
evidence from the likelihood analysis (equation \ref{bayes_like}).
The probability $p(\calM)$ is the prior probability of the model
in the absence of the data, for example from a previous
experiment. The evidence, the probability of getting the data
given the model for the system, is found by marginalizing over all
cosmological parameters in the model,
 \be
    E(\D|\calM) = p(\D|\calM) = \int d^{N_p} \!\theta
    \, L(\D|\thetab,\calM)
    p(\thetab|\calM).
 \ee
This can be estimated numerically using thermodynamic integration
(Slosar et al., 2003; Beltran et al., 2005), a variant of MCMC, or
by nested sampling (Skilling, 2004; applied to cosmology by
Bassett et al., 2004 and Mukherjee et al., 2006) or VEGAS, a
multi-dimensional integrator developed in particle physics
(Lepage, 1978) and applied in cosmology by Serra et al. (2007).
Alternative, approximate methods are the Savage-Dickey ratio for
nested models (Trotta, 2007), and the Bayesian Information
Criterion (BIC; Schwarz, 1987). When combining independent
dataset, parameter estimation only requires the addition of the
log-likelihoods, but the Bayesian evidence must be re-evaluated by
marginalization over the product of the posteriori distributions.
For a large parameter-space the estimation of the evidence can be
highly CPU-intensive, and so analytic methods are desirable.

\subsubsection{The Laplace Approximation}

There is already a well-known analytic marginalization method
which uses the saddle-point, or Laplace, approximation (see e.g.,
MacKay, 2003; Trotta, 2008), where the likelihood is expanded
around the peak in parameter-space;
 \be
    \calL_{\rm Laplace} = \calL_{\rm max} + \half \Delta \theta_i \Delta
        \theta_j \calL_{ij}
  \ee
where $\calL_{\rm max}$ is evaluated at the maximum of the
likelihood function in the full parameter space, and $\Delta
\thetab=\thetab - \thetab_{\rm max}$. With a flat prior,
$p(\theta|\calM) = 1/V_\theta $ where $V_\theta$ is the prior
volume of parameter space, we can carrying out the Gaussian
integration to find
 \be
    \calL_{\rm Laplace} = \calL_{\rm max} + 2 \ln ( V_\theta \sqrt{\det
        F_{ij}}).
        \label{laplace}
 \ee
The last term is again the \textit{Occam factor}, the ratio of the
prior (non-zero) volume of parameter-space to the effective
posterior volume measured by the parameter covariance matrix,
$\lgl \Delta \theta_i \Delta\theta_j \rgl=F^{-1}_{ij}$.

A severe limitation of the Laplace approximation is that the value
of $\calL_{\rm max}$ is evaluated at the maximum likelihood point
in parameter-space,
 \be
    \calL_{\rm max} = \calL(\theta_{\rm max}|\D,\calM),
 \ee
which depends on the data. Hence to evaluate it we must first find
the maximum likelihood for each model. To circumvent this,
embedded or nested models have been considered, where the relative
evidence between the evidence in one parameter-space can be
compared with that of a lower-dimensional parameter-space (see
e.g., Heavens, Kitching \& Verde, 2007).

\subsubsection{Analytic Evidence}

However, with analytic marginalization we now have a way to
estimate the maximum of the likelihood for an arbitrary dataset
and fixed fiducial parameter values (Section \ref{AM}). Expanding
the cosmological parameter space to second-order and
marginalizing, and this time keeping all terms, we find
 \be
    \calE  = \calL_0 - \half \calL_i
    \calL_{ij}^{-1}  \calL_j + \Tr \ln  V^{2/N_p}_\theta \calL_{ij}
    -N_p \ln 4 \pi.
    \label{am_evidence}
 \ee
where  $\calE \equiv -2 \ln E$ is the log-evidence.  This
expression is again then independent of the fiducial model used,
as we should expect after marginalization.

\subsubsection{Gaussian Likelihoods}

If the likelihood for the data is Gaussian and the parameters
appear in the mean, the evidence is
 \ba
 \label{eveq}
        \calE(\D|\calM) &=&
        \Delta \D \left(\C^{-1}- \C^{-1}\mub^t_i
        F_{ij}^{-1} \mub_j\C^{-1}\right)\Delta \D^t  \nn
        & +& \!\! \!\!\Tr \ln \C  +
        2 \ln ( V_\theta \sqrt{\det F_{ij}})-N_p \ln 2 \pi .
 \ea
The evidence is the probability based on the outcome of given
experiment. However we can also forecasting the evidence for
future experiments and ask what is the expected evidence, and even
what is the variance on a prediction of the evidence. Just like
the frequentist $\chi^2$-statistic, this will give us an
expectation of what the mean and range of values of evidence we
should expect from an experiment, give the uncertainty in the
data.

The expectation value of the Gaussian log-evidence is
 \be
    \lgl \calE \rgl = \nu + \Tr \ln \C +
    2\ln ( V_\theta \sqrt{\det F_{ij}}) -N_p \ln 2 \pi,
 \ee
where $\nu=N_D - N_p$ is the number of degrees of freedom, $N_D$
is the number of data points and $N_p$ is the number free
parameters. This is then just the $\chi^2$ number of degrees of
freedom, plus the normalization factor and the Occam factor. If we
were to ignore these terms, we see the Gaussian log-evidence,
$\calE$, has the same expectation value as the $\chi^2$-statistic.
If we further estimate the variance of the log-evidence we find
 \be
    \lgl \Delta \calE^2 \rgl = 2\nu,
 \ee
is just twice the number of degrees of freedom, as we might expect
for a Gaussian distribution. This  highlights the connection
between the evidence and the $\chi^2$-statistic, and shows that,
although they are asking different questions of the data, they
have a similar ``sensitivity''.

\subsubsection{Evidence for an arbitrary model}

In addition to calculating the evidence for the data, given the
maximum likelihood model also from the data, we can also ask what
is the probability that the measured data is drawn from an
arbitrary model, given an assumed set of ``true" parameter values,
$p(\D|\calM_t)$, and scatter in the possible data. We can
calculate this from
 \ba
 \label{eveq2}
        \calE(\D|\calM_t) &=&
        \Delta \D \C^{-1}\Delta \D^t
         + \Tr \ln \C  \\
         & & +
        2 \ln ( V_\theta \sqrt{\det F_{ij}}) - N_p \ln 2 \pi,
 \ea
where the likelihood peaks at the ``true" values, not the values
which best fit the data.  As an example, if the maximum likelihood
given the data peaks at a non-$\Lambda$CDM (non-standard model),
equation (\ref{eveq}) will yield the evidence for that model. But
instead if we assume that $\Lambda$CDM parameters is the ``true"
model, equation (\ref{eveq2}) will tells us the probability that
the data is drawn from this model. If this is very low, it is
unlikely the data is drawn from this model.

\subsubsection{The Occam Factor}

The final term in the evidence, the Occam factor, is often
problematic as it depends on the assumed prior volume of the
parameter space, which is not well-defined. While we can hope that
for good data the other terms in the evidence dominate over the
Occam factor, for poor data, this may not be the case. One
approach is to assume that the prior is set using the Fisher
matrix. We can let $V_\theta = a^{N_p} /\sqrt{\det F_{ij}}$, where
the constant of proportionality of order $a=10$ and $N_p$ is the
number of parameters. This factor becomes simply $2 N_p \ln a$,
and so this terms still gives more weight to models with fewer
parameters. The parameter $a$ becomes an adjustable parameter,
depending on how much weight one wants to give to the Occam
factor. A value of $a=10$ would seem to be fairly conservative.
Clearly this scheme can be extended for parameter which are highly
unconstrained.

We also note that our expression for the evidence will disfavour
models which have arbitrary un-constrained parameters. A common
concern in evidence calculations is that an extra parameter
entirely unconstrained by the data could be added that would
result in the disfavourment of the model only via the Occam
factor. We find that in such an unconstrained model the $\chi^2$
term becomes infinity because the Fisher matrix element for these
parameters is zero and hence the probability of such models is
zero.

\subsection{Model Selection}

\subsubsection{Model Selection: Bayes Factor}

A common approach to model selection is the use of the Bayes
factor (Kass \& Raferty, 1995), the ratio of pairs of models or
its logarithm,
 \be
    \calB_{AB} = -2 \ln B_{AB}
    =\calE(\D|\calM_A)-\calE(\D|\calM_B).
 \ee
This has the advantage that we do not need to consider the
normalization factor, $p(\D)$, in Bayes equation
(\ref{model_bayes}). Jeffery (1961) has proposed a qualitative
scale based on these ratios.

\subsubsection{Model Selection: Model-Space}

An alternative is to rank-order models by their evidence, with a
uniform prior, $p(\calM)=1/N_M$, where $N_M$ is the number of
models. Even though we do not expect to have a complete set of all
possible models, we can still normalize the set we have to
estimate the posterior probability for each model, $\calM_A$;
 \be
    p(\calM_A|\D) = \frac{p(\D|\calM_A)p(\calM_A)}{
    \sum_B^{N_M} p(\D|\calM_B)p(\calM_B)},
 \label{bayesnorm}
 \ee
where we consider independent models to form a countable set. By
this definition, uncountable sets of models contain models that
can be distinguished by a continuous parameter, which is then just
a model with a variable parameter i.e. we class a model as the set
of parameters, not a set of parameter values.

Even though the models may be incomplete, $p(\calM_A|\D)$ is an
upper limit on the true probability for each model with this
dataset. Adding any new model will only reduce the probability.
Since the prior is uniform, we expect a new model to appear at
random in the distribution. This scheme not only assesses
``goodness-of-fit'' to the data, but also the competitiveness of
models. If one model does well compared to other proposed models,
we rightly attach more belief to it. However, it does not prevent
a new model appearing with a higher evidence which would become
the best model. In this scheme, one would not necessarily truncate
or throw away models, since they contribute to the normalization
of the probabilities -- although if the contribution is negligible
it would seem sensible to drop outliers so the model-space is of a
manageable size.

\begin{figure*}
 \includegraphics[width=2.1\columnwidth]{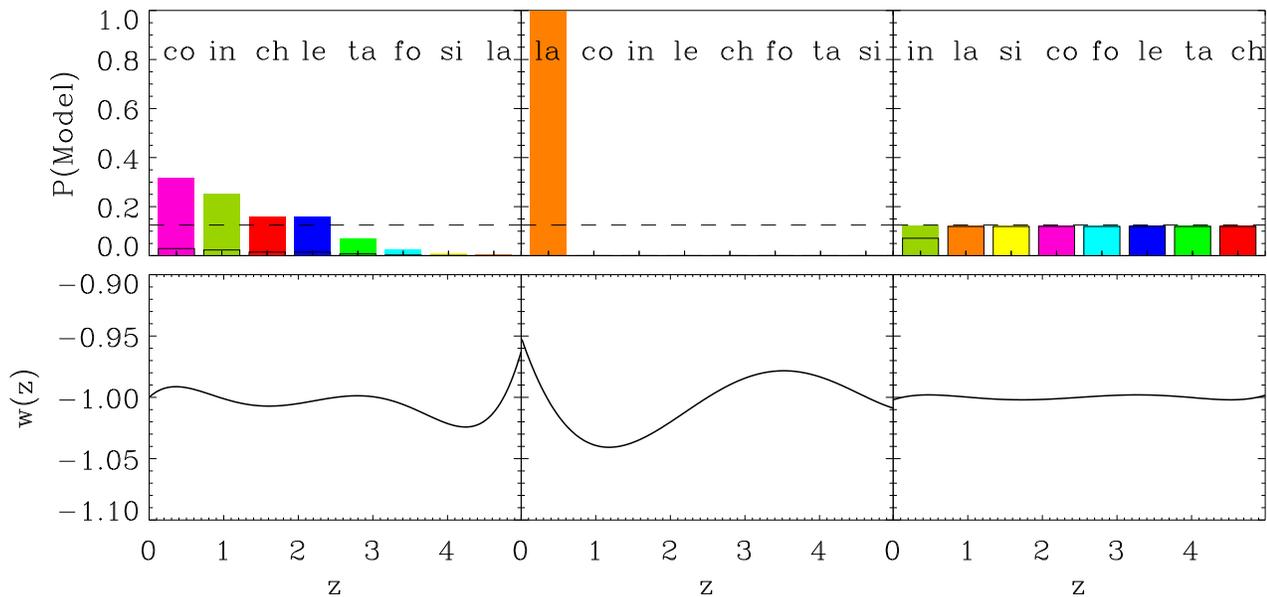}
 \vspace{-4.cm}
 \caption{\em A simple example of non-nested evidence analysis. The
   bottom row shows three $w(z)$ realizations, the top row shows the
   corresponding rank-ordered,
   non-nested evidence for each model on the left (using a Euclid weak
   lensing tomography experiment). The models are
   fo=Fourier (turquoise), ch=Chebyshev (red),
   la=Laguerre (orange), Le=Legendre (blue),
   in=Interpolation (dark green), ta=Taylor (light green),
   co=Cosine (purple) and si=Sine (yellow)
   (see Kitching \& Amara, 2009, for details). These represent the
   three possible classes of expected model space, a broad variance
   but with a favoured model; a highly favoured model; or a
   broad set of equally favoured models. In solid outlined bars we
   show the evidence that the data is drawn from a $\Lambda$CDM
   cosmology instead of the best fit values to the data.
   The dashed line show the flat model prior, $p(\calM)=1/N_M$.}
 \label{ev}
\end{figure*}

\subsubsection{Model Significance}

Even though the scheme outlined above puts an upper limit on the
absolute model probability it will still return the following
result: that if we only have one model, Bayes Theorem tells us we
must assign it a 100\% probability (since it is the only viable
model). Instead we could judge a model in relation to the prior we
assign it. To do this, we define a significance factor,
 \be
    \calS =\frac{p(\calM|\D)}{p(\calM)} =
    \frac{p(\D|\calM)}{p(\D)},
 \ee
where, by definition, $\calS \ge 1$, since we cannot lose
information by adding data. The evidence for any model is only
\emph{significant} if the ratio, $\calS$ of the evidence to the
prior for the model $\calM$ is much larger than unity. For
example, if we consider again the situation when we only have one
model the prior probability is $p(\calM)=1$, so that $\calS=1$,
and we have not learned anything about the absolute validity of
the model.

We can now estimate the number of models needed for any model to
be convincing in an absolute sense. For two models the uniform
prior for each model is $p(\calM_A)=1/2$, so the maximum
significance is $2$. While the Bayes factor between the two models
could `decisively'  favour one model over the other (odds of $\gs
1:100$ on Jefferys Scale), one could only be at most
`inconclusive' (odds of 1:2) that the model is correct. For
absolute confidence we need at least $3$ models for
comparison\footnote{Note the prior on the model is important here.
A flat prior of $1/N_M$ is only appropriate for equally credible
models. Including a vast array of non-credible models can be
countered by giving these a low-prior weighting.}. This argument
can be used to retrospectively understand the history of model
selection. For example, when given the choice of a Steady State
model over the Big Bang the later was clearly favoured due to a
large Bayes factor. However the absolute confidence in the Big
Bang could not be high since there were no alternative theories.
Indeed once Inflationary cosmologies appeared this new theory
became preferable.

If a new model is added to the model-space, the significance,
$\calS_A$ scales as
 \be
    \calS'_A = \frac{\calS_A  (N_M+1)}{N_M +
     \calS_A \frac{p({\small \D}|\calM_{\rm new})}{
     p({\small \D}|\calM_A)}}.
 \ee
If the new model has lower probability the significance scales as
$\calS'_A = \calS_A (N_M+1)/N_M$, while if it has much higher
probability it scales as $\calS'_A=
(N_M+1)p(\D|\calM_A)/p(\D|\calM_{\rm new})$.

\subsubsection{Dark Energy Model-Space}

In Figure \ref{ev} we show an example of how the evidence can be
used in practice, for the predicted evidence for a Euclid
(Refregier et al., 2006) weak lensing tomography experiment to
measure dark energy. In this example we have assumed a dark energy
equation of state, $w(z)$, as a function of redshift, $z$, which
we use to construct mock lensing data. We fit this data using
models that assume a cosmology with different $w(z)$ models. We
have chosen some non-nested basis set expansions for our $w(z)$
models these have a maximum order of $2$ (these phenomenological
models are described in Kitching \& Amara, 2009). For each $w(z)$
realization we rank-order the evidence for each model. In the
first example the Cosine model has the highest probability with
$0.4$ and the distribution in model space is Gaussian-like. In the
second example the Chebyshev model fits the data very well,
creating a spike in model space. In the third example there is no
model that favours the data over any other. These three example
represent the three broad classes of behaviour we can expect for
real data, where we hope for example 2 with a spike in
model-space. The variance in model-space is also an interesting
quantity, reflecting both the distinguishability of the models and
the quality of the data for model selection.

\section{Discussion}

We have presented new, analytic methods for Cosmological
Likelihood analysis to solve the ``many parameters" problem in
Cosmology. Our approach maximizes the likelihood with a
Pseudo-Newton Method, analytically marginalizes over nuisance
parameters in an arbitrary likelihood function, and analytically
marginalizes over cosmological parameters to project out one and
two-dimensions of parameter space to estimate marginalized errors
and covariance matrices. Parameters may have either flat or
Gaussian priors. Marginalizing over all parameters we derive an
analytic expression for the Bayesian evidence to select between
competing Cosmological models. The marginalized likelihood does
not degrade information about the remaining parameters, and  the
marginalized parameter information is preserved in the Fisher
Information matrix. The marginalized likelihood is also
independent of the fiducial model when the underlying likelihood
is exactly Gaussian.

We have applied our results to multivariate Gaussian likelihoods
for the data, where the marginalized parameters appearing in
either the mean of the data or its covariance matrix. An exact
result for a normalization nuisance parameter is found and applied
to the problem of estimation the matter density parameter,
$\Omega_m$, from galaxy power spectra, where the normalization,
which depends on the galaxy bias parameter, $b$, is marginalized
out. The analytic marginalization is found to be very close to
numerical marginalization. Analytic marginalization can also be
used to project parameter-space onto lower-dimensions to allow a
simple visualization of the full likelihood function.

We describe a semi-analytic marginalization method which could be
carried out by identifying Gaussian and non-Gaussian parameters
and treating them analytically and numerically, respectively, in
semi-analytic marginalization. An example is presented of a
3-parameter dark energy model with $(w_0,w_a, h)$, and again the
1-d analytically marginalized distribution is in very good
agreement with the numerical one. We extend this to an 8-parameter
model, where we highlight non-Gaussianity in the 2-d projected
distribution which is missed by the Fisher Matrix approximation.

Finally, we have also applied our analytic marginalization method
to find a closed expression for the Bayesian evidence and shown
its relation to the Laplace approximation. We discuss the case of
multivariate Gaussian-distributed datasets. We consider the Bayes
Factor, the ratio if the evidence of two models, and discuss the
properties of the full model-space posteriori distribution,
$p(\calM)$. We also introduce the significance of the model, the
degree by which the model evidence changes with respect to the
uniform prior. Finally we have illustrated our model selection
scheme on a set of non-nested dark energy models. Our method has
applications in Cosmological parameter estimation and model
selection, and many wider applications in the statistical analysis
of data .


\noindent{\em Acknowledgements:} We thank Andrew Liddle, John
Peacock, Alan Heavens, Fergus Simpson, Adam Amara, and Benjamin
Joachimi for much useful discussion. We also thank the DUEL
network (MRTN-CT-2006-036133) for supporting part of this work.
TDK is supported by STFC rolling grant number RA0888.



\section*{Appendix A: Gaussian Integration}

In this Appendix we derive equation (\ref{maglike}). Expanding the
log-likelihood to second order we find
 \be
  \calL = \calL_0 + \delta \psi_\alpha  \calL_\alpha
 +\half \delta \psi_\alpha \delta \psi_\beta
 \calL_{\alpha\beta} .
 \ee
By completing the square this can be rewritten as
 \be
    \calL = \calL_0 +
    \half \calL_{\alpha \beta}
    (\calL_\gamma \calL^{-1}_{\gamma \alpha}+ \delta \psi_\alpha)
    (\calL_\delta \calL^{-1}_{\delta \beta}+ \delta \psi_\beta)
    - \half \calL_\alpha \calL_{\alpha \beta}^{-1} \calL_\beta.
 \ee
Now writing the likelihood explicitly we find
 \be
    L = e^{-\half \calL_0 +
    \frac{1}{4} \calL_\alpha \calL_{\alpha \beta}^{-1} \calL_\beta
    -\frac{1}{4} \calL_{\alpha \beta}
    (\calL_\gamma \calL^{-1}_{\gamma \alpha}+ \delta \psi_\alpha)
    (\calL_\delta \calL^{-1}_{\delta \beta}+ \delta \psi_\beta)}
 \ee
Integrating over $\delta \psib$, and using the multivariate
Gaussian formula
 \be
        \int d^n x  \, e^{-\half x_i C_{ij}^{-1} x_j}=
        (2 \pi)^{n/2} \sqrt{\det \C},
 \ee
we find
 \be
    L = e^{-\half \calL_0 + \frac{1}{4} \calL_\alpha
    \calL_{\alpha \beta}^{-1}
    \calL_\beta}
    (2 \pi)^{N/2} \sqrt{\det 2 \calL^{-1}_{\alpha \beta}}.
 \ee
 Taking the log again we find
 \be
    \calL = \calL_0 - \half \calL_\alpha \calL_{\alpha \beta}^{-1}
    \calL_\beta  +  \ln \det \half \calL_{\alpha\beta}
    -N \ln 2 \pi.
 \ee
Using the identity $\ln \det M = \Tr \ln M$ yields equation
(\ref{maglike}).

\section*{Appendix B: Generating function}

The generating function of a distribution is
 \be
    \Phi(\J) = \lgl e^{i {\small \J . \delta \psib}}\rgl
    = \int d^M \psi \, e^{-{\small \calL}/2}
    \,e^{i {\small \J . \delta \psib}}
 \ee
which leads to the generating function of the likelihood;
 \be
  -2 \ln \Phi(J) = \calL_0 - \half (\calL_\alpha -2  i J_\alpha)
  \calL_{\alpha \beta}^{-1}
  (\calL_\beta -2 i J_\beta)
   + \Tr \ln \half \calL_{\alpha\beta}.
    \label{magchar}
 \ee
Taking the first derivative with respect to $i J_\alpha$ we find
the mean is
 \be
    \lgl \delta \psi_\alpha \rgl =
     \frac{\de \ln \Phi}{\de (i J_\alpha)} \Big|_{J=0}
     = - \calL_{\alpha \beta}^{-1}
    \calL_\beta(\thetab).
 \ee
For a Gaussian the mean is also at the peak, so this is a offset
between a fixed-point, $\psib_0$, where the likelihood is
evaluated and the peak. The second derivative yields the
covariance matrix
 \be
  \lgl \delta \psi_\alpha \delta \psi_\beta \rgl =
   \frac{\de^2 \ln \Phi}{\de (i J_\alpha) \de (i J_\beta)} \Big|_{J=0}
   = 2 \calL_{\alpha \beta}^{-1} .
  \ee
 Taking the ensemble average of the data we see
 \be
    \lgl \delta \psi_\alpha \delta \psi_\beta \rgl = F_{\alpha
    \beta}^{-1}
 \ee
as expected. Expanding $\thetab$ around its maximum-likelihood
value we find
 \be
    \lgl \delta \psi_\alpha \rgl =- \calL_{\alpha \beta}^{-1}
    \calL_{\beta i} \Delta \theta_i.
 \ee
Finally, inverting this we find the bias in cosmological
parameters, $\delta \thetab$, due to an offset in the nuisance
parameter is given by
 \be
        \delta \theta_i = - \calL_{i \alpha}^{-1}
        \calL_{\alpha\beta}\delta \psi_\beta.
 \ee
in agreement with the result of Taylor et al. (2007).

\label{lastpage}


\begin{thebibliography}{}


\bibitem{} Bassett B. A., Corasaniti P. S., Kunz M., 2004, Astrophys.
            J., 617, L1
\bibitem{} Beltran M., Garcia-Bellido J., Lesgourgues J., Liddle A. R.,
            Slosar A., 2005, Phys. Rev., D71, 063532
\bibitem{} Bretthorst G., 1988, in \textit{Bayesian Spectrum Analysis and
        Parameter Estimation}, Springer
\bibitem{} Bridle S.L., Crittenden R., Melchiorri A.,
            Hobson M.P., Kneissl R., Lasenby A.N., 2002, MNRAS, 335, 1193
\bibitem{} Eisenstein D., Hu W., 1997, ApJ, 511, 5
\bibitem{} Gamerman D., 1997, in \textit{Markov Chain Monte Carlo:
            Stochastic simulation for Bayesian inference}, Chapman and Hall.
\bibitem{} Gull S.F., 1989, in \textit{Maximum Entropy and
            Bayesian Methods}, Cambridge 1988, Ed. J. Skilling,
                p.511, Dordrecht: Kluwer
\bibitem{} Heavens A.F., Taylor A.N., 1995, MNRAS, 275, 483
\bibitem{} Heavens A.F., Kitching T.D., Taylor A.N., 2006, MNRAS, 373, 105
\bibitem{} Heavens A.F., Kitching T.D., Verde L., 2007, MNRAS, 380, 1029
\bibitem{} Jeffreys H., 1961, in \textit{Theory of probability}, 3rd edn., OUP
\bibitem{} Kaiser N., 1988, MNRAS, 231, 149
\bibitem{} Kass R.E., Raftery A.E., 1995, Bayes factors, J. Am. Stat. Assoc., 90, 773
\bibitem{} Kitching T., Amara A., 2009, arXiv:0905.3383
\bibitem{} Kitching T.,  Heavens A.F., Verde L., Serra P.,
            Melchiorri A., 2008, Phys.Rev.D, D77, 103008
\bibitem{}  Kosowsky A., Milosavljevic M., Jimenez R., 2002, Phys.Rev. D, 66, 063007
\bibitem{} Lepage G.P., 1987, J. Comput. Phys., 27, 192
\bibitem{} Lewis A., Bridle S., 2002, Phys. Rev. D 66, 103511
\bibitem{} Liddle A.R., 2009, Annual Reviews of Nuclear and Particle Science (ARNPS), vol 59
\bibitem{} MacKay D.J.C., 2003, in \textit{Information theory,
            inference, and learning
            algorithms}, Cambridge University Press, Cambridge
\bibitem{} Mukherjee P., Parkinson D., Corasaniti P.S., Liddle
            A.R., Kunz M., 2006, MNRAS, 369, 1725
\bibitem{}  Press W. H., Flannery B.P., Teukolsky S.A.,
            Vetterling W.T., 1989, in
            \textit{Numerical Recipes: The Art of Scientific Computing}, CUP
\bibitem{} Refregier A., et al., 2009, Exper.Astron, 23, 17
\bibitem{} Serra P., Heavens A., Melchiorri A., 2007, MNRAS, 379, 169
\bibitem{} Skilling J., 2004, in
            \textit{Baysian Inference and Maximum Entropy Methods in
                 Science and Engineering}, AIP Conference Proceedings,
                            Volume 735, p. 395
\bibitem{} Schwartz G., 1987, Ann. Statist., 5, 461
\bibitem{} Slosar A., et al., 2003, MNRAS, 341, L29
\bibitem{} Smith R., et al., 2003, MNRAS, 341, 1311
\bibitem{} Taylor A.N., Watts P., 2001, MNRAS, 328, 1027
\bibitem{} Taylor A.N., Kitching T., Bacon D., Heavens A.,
            2007, MNRAS. 374, 1377
\bibitem{} Tegmark M., Taylor A.N., Heavens A.F., 1997,
            Astrophys.J., 480, 22
\bibitem{} Tegmark M., 1997, Phys.Rev.D55, 5895
\bibitem{} Tegmark M., et al., 2004, Phys. Rev. D, 69, 103501
\bibitem{} Trotta R., 2007, MNRAS, 378, 72
\bibitem{} Trotta R., 2008, Contemp.Phys., 49, 71
\bibitem{} Woodbury M.A., 1950, Inverting modified matrices,
             Memorandum Rept. 42, 4, Statistical Research Group, Princeton
             University
\bibitem{} Zhang F., 2005, in \textit{The Schur Complement and
             its Applications},
            Springer

\end{thebibliography}
\end{document}